# Unleashing the potential of price-based congestion management schemes: a unifying approach to compare alternative models under multiple objectives

Ennio Cascetta [a], Marcello Montanino [b, *]

[a] Mercatorum University, Piazza Mattei, 10, 00186 Roma, IT
[b] University of Naples Federico II, Via Claudio, 21, 80125 Napoli, IT

[*] Corresponding Author: Ph.: +39 081 7683770

E-mail addresses:
ennio.cascetta@unimercatorum.it
marcello.montanino@unina.it

## ABSTRACT

A wide range of price-based congestion management schemes were proposed in the literature ranging from marginal cost road pricing to trip based multimodal pricing. The underlying models were formulated under different theoretical assumptions and with varying, and sometimes conflicting objectives. This paper presents a unifying framework under which different approaches can be compared based on their respective assumptions. The unifying modelling framework is referred to as *trip pricing* model – which extends path-differentiated pricing to multimodal paths, i.e., trips on whatever mode or combinations of modes, and generalizes road pricing schemes (which are shown to be a special case). By setting both positive (tolls) and negative (incentives) prices, revenue-neutral schemes are also shown to be special cases, with no a-priori assumption on which paths/modes to toll/incentivize. For model comparison, the pricing design problem is formulated in a multi-objective optimization framework, which combines traffic efficiency, environmental sustainability, users' acceptance, social and spatial equity, as pricing objectives. The trip pricing scheme is compared with traditional road pricing schemes, and with their revenue-neutral variants. First-best and second-best pricing schemes, designed in single- and multi-objective optimization frameworks, are compared on the Nguyen-Dupuis network. Results suggest a vast potential for multi-objective trip congestion pricing over single-objective road pricing. In addition, the application of both positive and negative prices is shown to significantly increase the expected acceptance of pricing schemes and to preserve efficiency and sustainability objectives. The results should promote the design of more effective pricing policies, i.e., set of pricing rules easier to communicate, based on now available technologies of passengers' tracking and pricing.

**Highlights:**

1. A multi-objective optimal design problem with five pricing objectives is specified
2. Congestion pricing models are formulated under stochastic user equilibrium
3. Two novel MoPs are proposed to quantify users' acceptance and equity impacts
4. Single/multi-objective trip and road pricing model performances are compared
5. Results show a large potential of more sophisticated models over traditional ones

## 1. INTRODUCTION

Congestion has long been recognized as an inherent regulator of traffic networks, albeit a very inefficient one. By directly influencing travellers' choice, road pricing has been proposed as a solution to push transportation networks towards more (traffic) efficient equilibrium conditions (Vickrey, 1963). Since then, tremendous technological improvements have become available in terms of economic transactions in the transportation market, and price-based policies have become much more common in all commodity markets, including revenue management for air and high-speed rail systems.

Despite its undiscussed potential benefits and the technological advances, from a practical viewpoint congestion road pricing has faced a number of problems as an effective policy tool. Existing implementations have been limited (mostly cordon-based in central urban areas , which only partially unleash the true potential of pricing) and price levels have been mostly conceived to meet public acceptance, rather than designed to be optimal (e.g., Lehe, 2019). In even more cases, instead, road pricing proposals have been outright rejected mainly due to the lack of consensus (see examples in Gu et al., 2018). In fact, having car drivers pay for 'something' that has been traditionally free is a hard sell, even if congestion decreases and public transportation performances improve. On the contrary, pricing on newly built roads is very common and well accepted since 'additions' to the previous system are willingly paid for. Also in these cases, though, flat prices are applied thus wasting the opportunity to reduce congestion and increase sustainability of the road system, not to mention the multimodal system.

Negative experiences with congestion pricing have thus prompted theoretical developments of traditional pricing models towards equity and acceptance issues, as well as in the verification of pricing scheme performances in multimodal networks (e.g., Yang and Zang, 2002; Yin and Yang, 2004; Hamdouch et al., 2007; Liu et al., 2009; Song et al., 2009; Nie and Liu, 2010; Guo and Yang, 2010; Wu et al., 2011; Wu et al., 2012; Yu et al., 2017). A variety of price-based models have been developed in the last thirty years to address the optimal pricing design problem. The resulting literature has become very complex, with many alternative formulations, grounded on very different (and often contrasting) assumptions, and tackling the design problem from different angles, e.g., in relation to design objectives.

Therefore, the first contribution of this work is to provide a unifying framework, built upon a common theoretical background, to revisit the major model developments from the literature, and to identify open issues (Section 2). In fact, a thorough review of the literature revealed that most pricing models have been conceived in a link-based setting, developed under the assumption of deterministic traffic assignment, and adopted traffic efficiency/social welfare as the main design objective. By means of the proposed framework, link-based models were proven inadequate to effectively take into account the spatial heterogeneity of travel costs, thus contributing to increase spatial inequity. Moreover, pricing schemes designed in a deterministic traffic assignment context were shown to be incapable to reach system optimum in presence of stochastic user choice behaviour, even in very simple networks. Eventually, the customary design technique based on Marginal Social Cost pricing theory was shown to return prices which are far from the minimum toll guaranteeing system optimum conditions, thus decreasing users' acceptance of so-designed schemes.

To address the above limitations, a unifying modelling framework is first proposed to extend and generalize the models available in the literature (Section 3), and to allow a comparison of their performances. The generalized model applies pricing to paths, representing possible trips on *whatever mode or combinations of modes* are simulated in the transportation network, rather than on (car-only) links. For this reason, it is referred to as *trip pricing* model in contrast to road pricing model (which is shown to be a special case). The trip pricing model extends existing design approaches of path-based congestion pricing modelling using a multimodal and multiclass, *stochastic* user equilibrium assignment, while most model proposed in the literature derived from Wardrops' deterministic equilibrium. In addition, the model allows the application of both *positive (tolls) and negative*

*(incentives) prices* enabling revenue-neutral schemes at the path level, *without* setting constraints on what paths/modes can be tolled/incentivized, which is a novelty in the field. Finally, the pricing design problem is formulated in a *multi-objective optimization* framework, thus overcoming the limitations of existing approaches (such as design with objectives as constraints). The framework combines all *five main pricing objectives* in a scalarized objective function, i.e., traffic efficiency, environmental sustainability, users' acceptance, social equity, and spatial equity. In particular, *two novel measures of performance*, based on the satisfaction variable derived from choices among travel alternatives, were proposed to quantify users' acceptance and social/spatial equity.

It is worth noting that the proposed price-based modelling framework does not include tradable credit schemes (see e.g., Yang and Wang, 2011), since such schemes require the establishment of an external market to buy and sell mobility credits. The extension of the proposed framework to tradable credit schemes is beyond the scope of this work.

Basing on the proposed generalized modelling framework, the second contribution of the paper is the comparison among performances of trip pricing schemes with traditional road pricing schemes, and with their revenue-neutral variants (Section 4). Models were compared in the Nguyen-Dupuis fictional network ( often adopted to benchmark congestion pricing schemes in the literature), in two scenarios: a car-only network scenario with fixed demand, and a multimodal network scenario with mode choice elastic demand. Both first-best and second-best pricing schemes were tested. Numerical results allowed to i) verify experimentally theoretical properties of the proposed trip pricing model, and ii) quantify the impact on pricing performances of different model variants. Clearly, the absolute model performances measured in this experiment are not intended to be generally transferable to different network configurations.

Eventually, a technology-ready solution to implement the proposed model is discussed, together with potential privacy concerns (Section 5). Conclusions are reported in Section 6.

## 2. A REVIEW OF EXISTING PRICE-BASED APPROACHES

Inspired by the seminal work of the English economist Pigou (Pigou, 1920), the bulk of works on congestion pricing start flourishing in the transportation field from the nineties of the last century. Through Scopus®, we have implemented a comprehensive literature search on "congestion pricing" and scrutinized over 350 papers which either were published on journals that are widely recognized by transportation policy and modelling researchers as premier venues for dissemination or have been well-received in the literature. A selection of 40 papers, which depict a wide range of congestion pricing models moving from very different assumptions (e.g., first-best/second-best models, credit-based models, Pareto-improving models, etc.), have been classified in Table 1 according to three shared dimensions:

- the network element which pricing scheme is applied to, i.e., links or paths;
- the User Equilibrium (UE) assignment model adopted, i.e., Deterministic (DUE) or Stochastic (SUE);
- the number and type of objectives/goals which are considered in the congestion pricing design problem.

According to Table 1, the vast majority of papers in the field literature concentrated efforts on link-based congestion pricing models, i.e., models aiming at designing tolls either for each link (first-best pricing) or for a subset of links (second-best pricing), rather than on path-based models. This choice was motivated by i) the lower complexity of link-based design problems than path-based ones, as the number of links is usually much lower than the number of paths in a network, and ii) the higher level of readiness of technology that support link-based implementations, such as automatic number plate recognition, short-range communication, and on-board unit systems (for a review, see de Palma and Lindsey, 2011).

Nowadays, the availability of increasingly higher computing power which allows solving complex network problems in feasible time, on one hand, and of pervasive connectivity through smartphones, on the other hand, is key to enable path-based congestion pricing applications (see Section 4 for a discussion on practical implementations). Therefore, in the near future an increasingly higher research effort will be devoted to the development of path-based models for congestion pricing. Moreover, these models can overcome some limitations of link-based ones, such as the ability of thoroughly accounting for spatial equity effects of congestion pricing (see Section 2.2), thus allowing for an effective multi-objective design of the pricing scheme.

| Pricing scheme | Traffic assign. | Single objective | Multiple objectives | | | | |
| --- | --- | --- | --- | --- | --- | --- | --- |
| | | *Eff* | *Eff+Env* | *Eff+Acpt* | *Eff+Equ* | *Eff+Env+ Equ* | *Eff+Env+ Acpt+Equ* |
| *Link-based* | *DUE* | Verhoef et al. (1996), Yang and Huang (1998), Liu and McDonald (1999), Verhoef (2002), Yang et al. (2004), Zhang and Yang (2004), Hamdouch et al. (2007), Yin and Lawphongpanich (2009), Lou et al. (2010), Jeff Ban et al. (2013) | Johansson (1997), Johansson (2006), Yin and Lawphongpanich (2006), Chen and Yang (2012), Coria and Zhang (2017) | Eliasson (2001)[2], Song et al. (2009)[2], Liu et al. (2009)[2], Lawphongpanich and Yin (2010)[2], Guo and Yang (2010)[2], Nie and Liu (2010)[2], Liu and Nie (2017) | Yang and Zang (2002)[1], Yin and Yang (2004)[1], Sumalee et al. (2009), Yu et al. (2017)[4] | | |
| | *SUE* | Yang (1999)[6], Cascetta (2001), **This paper** | Li et al. (2012) | Kalmanje and Kockelman (2004)[3], Kockelman and Kalmanje (2005)[3], Zhao and Kockelman (2006), Wu et al. (2011), Liu et al. (2014) | Wu et al. (2012), Zheng and Geroliminis (2020) | | ***This paper*** |
| *Path-based* | *DUE* | Zangui et al. (2013), Zangui et al. (2015a), Ren et al. (2020) | | | | Sun et al. (2016)[5] | |
| | *SUE* | ***This paper*** | | *Liu et al. (2017)* | | | ***This paper*** |

*Eff = traffic efficiency/social welfare, Env = environment, Acpt = users' acceptance, Equ = equity (social/spatial)*

[1] *Equity is treated as a constraint.*
[2] *User acceptance is treated as a constraint, i.e., Pareto-improving constraint.*
[3] *User acceptance is indirectly taken into account through a credit-based scheme with an initial credit budget and subsidies.*
[4] *Equity is indirectly taken into account by adding a constraint of non-tollable alternative paths.*
[5] *Environment and equity are treated as constraints.*
[6] *Traffic efficiency is measured through the total generalized cost, rather than the total travel time. See Section 2.3.1 for a discussion.*

**Table 1 – Literature review and paper contributions.**

In particular, the type of equilibrium model adopted in the pricing design problem is the second classification dimension of Table 1. Most of the literature focused on DUE models, which directly follow the Wardropian theory. Such approach was pursued mainly for its consistency with the Pigouvian theory (see Section 2.1), and for theoretical analysis of proposed models, e.g., for the sake of proving analytical properties of pricing models like the existence and unicity of second-best,

Pareto-improving or revenue-neutral solutions (e.g., Verhoef, 1996; Liu and McDonald, 1999; Eliasson, 2001; Yin and Lawphongpanich, 2009).

It comes unsaid that the above-mentioned approach is less suitable to capture users' travel behaviours and may limit the validity of theoretical findings once such pricing models are plugged-in into Random Utility (RU) demand modelling frameworks. In addition, as mentioned in Liu et al. (2014) and Liu et al. (2017), considering SUE, rather than DUE, in the pricing design problem is more consistent with users' choice behaviours.

The third dimension in Table 1 refers to the objectives of the pricing design problem. If classical models have been developed with a single objective (efficiency-oriented pricing models), models that account simultaneously also for all main objectives, i.e., efficiency, environment, users' acceptance, and equity, have never been evaluated in the field literature so far.

In the next sub-sections, we discussed approaches from Table 1 over more specific topics. The notation applied throughout the paper is summarized in Table 2.

## 2.1. Marginal Social Cost pricing vs. UE with pricing

In this section, a discussion about the classical theory of Marginal Social Cost (MSC) pricing and alternative pricing models is given from a path-based perspective.

Given a network $G(N, A)$, the set of all feasible path flow vectors, H, is given by:

$$H = \left\{ h \in \mathbb{R}_+^{|K|} \mid d = \mathbf{1}^T h \right\} \tag{1}$$

Under the assumptions of deterministic path choice behaviour and static traffic assignment, Wardrop's User Equilibrium (UE) and System Optimum (SO) conditions can be formulated as follows:

$$(g^{ad}(h^{UE}) + g^{nad})^T (h - h^{UE}) \geq 0, \forall h \in H \tag{2}$$

$$h^{SO} = \arg \min_{h \in H} [(g^{ad}(h) + g^{nad})^T h] \tag{3}$$

where the additive path cost function, $g^{ad}$, is a linear combination of the link cost function, $c$, i.e., $g^{ad} = \Delta^T c(f)$. It is worth noting that, in Eqs. 2 and 3, non-additive path costs, $g^{nad}$, i.e., path costs which do not result from the sum of link costs, can be present also without applying any congestion pricing scheme (they are representative of e.g., public transport fares, motorway tolls, etc.).

Conditions in Eq. 2 and 3 are formulated in terms of path flows. Equivalent conditions on link flows can be easily derived by applying the following supply model equation:

$$f = \Delta h, \forall h \in H \rightarrow F = \left\{ f \in \mathbb{R}_+^{|A|} \mid f = \Delta h, \ \forall h \in H \right\} \tag{4}$$

In congested networks, UE and SO conditions yield two different flow distributions (see the example on a two-link network in Appendix A).

In case of separable link cost functions, Eq. 3 implies that:

$$(g^{ad}(h^{SO}) + g^{ad}(h^{SO})' h^{SO})^T (h - h^{SO}) \geq 0, \forall h \in H \tag{5}$$

By comparing the SO condition in Eq. 5 with the UE variational inequality in Eq. 2, one can easily recognize that UE and SO conditions coincide if non-additive path costs are equal to the marginal external path cost at SO, i.e., $g^{nad} = g^{ad}(h^{SO})' h^{SO}$. Such equivalence is at the basis of MSC pricing theory, which is a well-known technique applied to design congestion pricing tolls. In fact, if MSC tolls $\pi^{MSC} = g^{ad}(h^{SO})' h^{SO} = c(f^{SO})' f^{SO} = \gamma^{MSC}$ are added to path/link cost functions, UE with MSC pricing yields a SO flow distribution.

| Symbol | Description |
|---|---|
| W | Set of OD pairs, with cardinality $\|W\|$ |
| Q | Set of users' classes, with cardinality $\|Q\|$ |
| A | Set of links, with cardinality $\|A\|$ |
| K | Set of paths, with cardinality $\|K\|$ |
| M | Set of modes, with cardinality $\|M\|$ |
| $K_{m,w}$ | Set of paths of mode $m$ between OD pair $w$, where $m \in M$ and $w \in W$, with cardinality $\|K_{m,w}\|$ |
| $K_w$ | Set of paths between OD pair $w$, where $w \in W$, with cardinality $\|K_w\|$ |
| $\Delta^{m,w}$ | Link-path incidence matrix on mode $m$ between OD pair $w$, where $m \in M$ and $w \in W$, with dimensions $\|A\| \times \|K_{m,w}\|$ |
| $\Delta$ | Link-path incidence matrix, with dimensions $\|A\| \times \|K\|$, obtained by horizontal concatenation of $\Delta^{m,w}$ |
| $f^{q,m}$ | Vector of vehicle link flows of users of class $q$ on mode $m$, where $q \in Q$ and $m \in M$, with dimensions $\|A\| \times 1$, such that $f_i^{q,m} = 0, \forall i : \Delta_{i,:}^{m,w} = 0 \; \forall w$ |
| $f$ | Vector of vehicle link flows |
| $h^{q,m,w}$ | Vector of vehicle path flows of users of class $q$ on mode $m$ between OD pair $w$, where $q \in Q$, $m \in M$ and $w \in W$, with dimensions $\|K\| \times 1$, such that $h_k^{q,m,w} = 0, \forall k \notin K_{m,w}$ |
| $h$ | Vector of vehicle path flows, with dimension $\|K\| \times 1$, such that $h = \sum_{q \in Q} \sum_{m \in M} \sum_{w \in W} h^{q,m,w}$ |
| $d$ | Vector of passenger demand flows |
| $\psi^q$ | Vector of proportions of users of class $q$, where $q \in Q$ |
| $d^q$ | Vector of passenger demand flows of users of class $q$, where $q \in Q$, such that $d_w{}^q = \psi^q d_w, \forall w \in W$ |
| $c^q$ | Vector of generalized link costs for users of class $q$, where $q \in Q$, with dimensions $\|A\| \times 1$ |
| $c$ | Vector of generalized link costs, with dimensions $\|A\| \times 1$ |
| $g^{ad,q,m,w}$ | Vector of generalized additive path costs for users of class $q$ on mode $m$ between OD pair $w$, where $q \in Q$, $m \in M$ $w \in W$, with dimensions $\|K_{m,w}\| \times 1$ |
| $g^{nad,q,m,w}$ | Vector of generalized non-additive path costs for users of class $q$ on mode $m$ between OD pair $w$, where $q \in Q$, $m \in M$ $w \in W$, with dimensions $\|K_{m,w}\| \times 1$ |
| $g^{q,m,w}$ | Vector of generalized path costs for users of class $q$ on mode $m$ between OD pair $w$, where $q \in Q$, $m \in M$ $w \in W$, with dimensions $\|K_{m,w}\| \times 1$ |
| $g^{ad,q}$ | Vector of generalized additive path costs for users of class $q$, where $q \in Q$, with dimensions $\|K\| \times 1$ |
| $g^{nad,q}$ | Vector of generalized non-additive path costs for users of class $q$, where $q \in Q$, with dimensions $\|K\| \times 1$ |
| $g^q$ | Vector of generalized path costs for users of class $q$, where $q \in Q$, with dimensions $\|K\| \times 1$ |
| $l$ | Vector of link lengths, with dimension $\|A\| \times 1$ |
| $v$ | Vector of link speeds, with dimension $\|A\| \times 1$ |
| $cap$ | Vector of link capacities, with dimension $\|A\| \times 1$ |
| $TT(\cdot)$ | Link travel time function |
| $WT(\cdot)$ | Link waiting time function |
| $MC^{q,m}(\cdot)$ | Link energy consumption function |
| $\eta^{q,m}$ | Passenger occupancy factor of users of class $q$ per unit of vehicle of mode $m$, where $m \in M$ |
| $\zeta^{q,m}$ | Unit energy cost for users of class $q$ on mode $m$, with dimension $\|A\| \times 1$ |
| $VOT^q$ | Value of Travel time of users of class $q$, where $q \in Q$ |
| $VOWT^q$ | Value of Waiting time of users of class $q$, where $q \in Q$ |
| $s(\cdot)$ | User's satisfaction function |
| $s_w^{q'}$ | Satisfaction of users of class $q$, on OD pair $w$, per unit of average path length, where $q \in Q$ and $w \in W$ |
| $\theta^M$ | Mode choice dispersion parameter of logit model |
| $\theta^K$ | Path choice dispersion parameter of logit model |
| $p^{M\|q,w}$ | Vector of mode choice probabilities for users of class $q$ between OD pair $w$, where $q \in Q$ and $w \in W$, with dimensions $\|M\| \times 1$, such that $p_m^{M\|q,w} = 0, \forall m \in M : \|K_{m,w}\| = 0$ |
| $p^{K_{m,w}\|q}$ | Vector of path choice conditional probabilities for users of class $q$ on mode $m$ between OD pair $w$, where $q \in Q$, $m \in M$ and $w \in W$, with dimensions $\|K\| \times 1$, such that $p_k^{K_{m,w}\|q} = 0, \forall k \notin K_{m,w}$ |
| $p^{q,m,w}$ | Vector of path choice probabilities for users of class $q$ between OD pair $w$, where $q \in Q$ and $w \in W$, with dimensions $\|K\| \times 1$, such that $p_k^{q,m,w} = 0, \forall k \notin K_w$ |
| $\gamma$ | Vector of anonymous link prices |
| $\gamma^u$ | Vector of anonymous link prices per unit of length |
| $\pi$ | Vector of anonymous path prices |
| $\pi^u$ | Vector of anonymous path prices per unit of length |

Sets are indicated with uppercase letters, matrices and vectors with italic uppercase and lowercase letters, respectively. The $i$-th element of a vector $x$ is indicated with $x_i$, while the $(I, j)$-th element of a matrix $X$ is indicated with $X_{i,j}$.

## Table 2 – Notations.

The MSC toll vector, however, is just one of the possible toll vectors fulfilling the SO condition in terms of traffic efficiency/social welfare. More importantly, MSC tolls, i.e., $\gamma^{MSC}$, do not correspond to the minimum toll allowing to reach the SO condition, therefore hindering the users' acceptance of such pricing scheme (see an example in Appendix A with positive and negative tolls leading to SO, while guaranteeing a lower generalized user cost than UE, i.e., Pareto-improving pricing, Section 2.2, and a zero net revenue for the pricing operator, i.e., revenue-neutral pricing, Section 2.3). Explicitly allowing negative tolls only on public transport link/paths, has been explored in the literature. In this work, we evaluated the impact, on multiple pricing objectives, of setting no a-priori constraints on what paths/modes can be tolled/incentivized.

Schemes inspired to the MSC principle have been designed targeting at the maximization of traffic efficiency/social welfare, i.e., targeting at time/cost SO. However, as a transportation demand management policy, congestion pricing may target also other objectives, the main ones being emissions reduction (e.g., Johansson, 1997; Johansson, 2006; May and Milne, 2000; Yin and Lawphongpanich, 2006; Chen and Yang, 2012; Li et al., 2012; Coria and Zhang, 2017), increase of users' acceptance and equity (e.g., Yang and Zang, 2002; Yin and Yang, 2004; Sumalee et al., 2009; Wu et al., 2012; Zheng and Geroliminis, 2020). The formulation proposed here encompasses all mentioned objectives (i.e., traffic efficiency, environment, users' acceptance, and equity) in a generalized multi-objective pricing design model.

## 2.2. Pareto-improving pricing, credit-based schemes, and equity

Social and spatial inequity[1] produced by MSC pricing scheme – and by any scheme aiming at maximizing only traffic efficiency or social welfare – has been identified as a main factor hindering public acceptance (e.g., Rouwendal and Verhoef, 2006; Eliasson and Mattsson, 2006). Inequity of congestion pricing schemes originates from the (high) variability of perceived generalized travel cost across users' classes (social/vertical inequity) and across OD pairs (spatial/horizontal inequity).

Two main variations of the pricing model have been advocated in the field literature to account for inequity effects provoked by congestion pricing, and specifically: *i)* Pareto-improving schemes and *ii)* credit-based schemes. The objective of Pareto-improving schemes is to make no user class, over no OD pair, worse off compared to UE conditions (Song et al., 2009). It is achieved by guaranteeing that, for each users' class $q$, each generalized link cost in the pricing scenario, i.e., at $f^{SO}$, is not greater than corresponding cost in the non-pricing scenario, i.e., at $f^{UE}$. Formally, the Pareto-improving condition is formulated as follows:

$$c^q(f^{SO}) \leq c^q(f^{UE}), \forall q \in \mathbb{Q} \tag{6}$$

Eq. 6 is either applied as an additional constraint in the pricing scheme design problem (e.g., Song et al., 2009; Liu et al., 2009; Lawphongpanich, and Yin, 2010), or as objective function in the revenue-redistribution scheme design problem (e.g., Eliasson, 2001; Guo and Yang, 2010; Nie and Liu, 2010). However, lower travel costs on each link for each user class, relative to the UE condition, do not imply a lower *variability* of costs across classes (which would imply a higher equity, according to Ecola and Light, 2010). For instance, a reduction of the average travel costs on each network link for every user may leave unchanged the difference of travel costs among users. In other words, the Pareto-improving condition does not necessarily make the designed scheme more socially equitable. Moreover, it is worth highlighting that Pareto-improving schemes have been formulated in the field literature at the link-level only. Therefore, they are inherently unable to effectively capture spatial inequity effects. In fact, by pricing each link, the travel costs of each OD pairs cannot be controlled individually (as a link may belong to paths serving multiple ODs).

---

[1] These two types of equity have been also referred to as "vertical" and "horizontal" equity in the field literature (e.g., Ecola and Light, 2010).

In credit-based schemes (e.g., Kalmanje and Kockelman, 2004; Kockelman and Kalmanje, 2005; Gulipalli and Kockelman, 2008), instead, an initial credit endowment provided to each user is expected to alleviate the unequitable welfare effects of congestion pricing. However, link tolls in credit-based schemes are still designed through MSC pricing or Pareto-improving pricing/refunding schemes. Therefore, the initial credit endowment aims at making the policy tool more acceptable, rather than more socially equitable (i.e., travel costs less dispersed among user classes). In addition, such endowment is usually anonymous, i.e., not differentiated by user classes, and cannot be differentiated by OD trip (credits usage is not spatially restricted), thus making it uncapable to deal with spatial inequity.

In contrast to previous schemes, few works have proposed pricing models able to effectively account for equity concerns, according to the definition provided above. In these works, the level of equity is quantified through the Gini coefficient of the user generalized travel cost (Yang and Zang, 2002; Yin and Yang 2004; Sumalee et al. 2009; Wu et al., 2012; Sun et al., 2016), or its variance (Zheng and Geroliminis, 2020). However, the most frequently applied approach in the literature has been an equity-constrained single-objective optimization aiming at maximizing traffic efficiency. Only few studies attempted to design tolls by solving a multi-objective problem through a weighted Generalized Least Square (w-GLS) method with traffic efficiency and equity as objective function components (e.g., Wu et al. 2012; Zheng and Geroliminis, 2020). In these models, however, social and spatial equity components are not usually accounted simultaneously, rather only one of the two types is included in the model (social equity in Wu et al., 2012; and spatial equity in Zheng and Geroliminis, 2020).

## 2.3.  Revenue-redistribution vs. revenue-neutral schemes

Revenue redistribution has been proposed as a mean to increase public acceptance (see Section 2.2), by applying the concept of Pareto-improving condition in Eq. 6. Several pricing and redistribution schemes have been proposed in the literature, at the link-level (e.g., Small, 1992; Eliasson, 2001; Yang and Guo, 2005; Guo and Yang, 2010). In the vast majority of the proposed schemes, however, pricing and revenue redistribution are considered as consecutive steps. Therefore, the redistribution scheme may result in equilibrium flows in day-to-day dynamics which are different from those applied in pricing scheme design (e.g., with MSC). Moreover, users may dynamically change their mode/route choice behaviour depending on the amount of distributed refunds, which increases system instability.

In contrast to revenue redistribution schemes, revenue-neutral schemes allow for the simultaneous design of both positive prices (tolls) and negative prices (incentives), so that the resulting priced UE flows produced a zero net revenue for the pricing operator.

Such schemes, however, have been largely under-researched in the field literature. In the available studies, revenue-neutral models have been proposed at the link-level only. Moreover, either prices on transit links only were allowed to be negative (Liu et al., 2009; Zheng et al., 2016), or, if no a-priori assumption is made on links to be tolled or incentivized, proposed models have been developed and tested only on mono-modal networks (Bernstein, 1993; Adler and Cetin, 2001).

## 3.  GENERALIZED MULTI-OBJECTIVE PRICING MODELS

### 3.1.  Modelling assumptions

The proposed modelling framework is based on basic assumptions whose significance and limitations are discussed below.

- *Pricing model*
  In this work, a path-based pricing model is proposed, i.e., prices are differentiated by path. As shown in Section 3.2, link-based models are formulated as a special case of path-based ones. Since prices are designed for each path, explicit (and exhaustive) path enumeration would be necessary. This approach may generate a large number of paths that share many links and are correlated in their perceived utilities (see next point on "Demand model"). Furthermore, the computational complexity of explicitly enumerating all the paths in a network may require further assumptions to enable the application of the proposed model to real-world network. In particular, path prices may be designed exclusively for a selection of paths which are followed by users (being the collection of users preferences performed through the same technology that enable the application of the pricing scheme, see Section 5).

Concerning path prices, both positive prices, i.e., tolls, and negative prices, i.e., incentives, are considered, without any a-priori assumption on paths that are allowed to be tolled/incentivized or not. This choice allows a wider range of solutions with respect to those explored in the previous literature (see Section 2.3). Moreover, a revenue-neutral constraint is introduced in order to guarantee self-financing scheme.

Prices are assumed to be not differentiated with respect to users' socio-economic factors, i.e., prices are anonymous with respect to users' classes, i.e., all users pay the same price, regardless of their class. Clearly, if this assumption were relaxed, i.e., user class-specific prices are designed, the social equity objective could be further improved.

- *Pricing design problem*
  The pricing design problem is formulated as a w-GLS multi-objective optimization problem with traffic efficiency, environment, users' acceptance, social and spatial equity as objectives (see Section 3.2).

- *Demand model*
  The proposed model builds upon RU models to describe travellers' mode and path choice behaviour. Specifically, mode and path choices are modelled through a single-level hierarchical model. For the sake of simplicity, hyperpath choice behaviour is not explicitly modelled in this work (though an extension of the proposed model is straightforward).

  Taste heterogeneity across users' classes is modelled with a discrete distribution of reciprocal substitution coefficients (including VOT) in the systematic utility function of mode/path choice models.

  It is worth noting that the trip frequency, destination, and departure time choice models are not considered in this work.

  The system-wide demand elasticity is ignored, i.e., destination choices are not influenced by cost variations, but modal elasticity is considered, through mode choice modelling. If system-wide demand were also elastic, prices would affect demand as well as the modal split. Therefore, negative prices on paths of certain ODs, i.e., incentives, may capture demand from other ODs, or even generate new demand (if trip production were considered). Such higher-level effect is not considered in this paper, as in most papers dealing with congestion pricing.

- *Supply model*
  The generalized link cost function is assumed to be a linear combination of a link travel time function (e.g., BPR) and a link-based specific energy/fuel consumption function. Both functions are class $C^2$, convex, and separable. On road links, the travel time function is strictly increasing with link flow (BPR function), while the specific energy consumption function is not. In fact, it is well-acknowledged that specific (average) energy consumptions of real-world vehicles exhibit a quadratic trend (e.g., Fiori et al., 2019; Fiori et al., 2021a). However, it is worth noting that the rate of change of the specific energy consumption at high speeds (i.e., at low travel time, that is at low link flow) is much lower than the one at low speeds (i.e., at high travel time, that is at high link flow). Therefore, in case of urban congestion pricing applications (average speed lower than 80 km/h), also the energy consumption function can be considered as strictly increasing, thus preserving the unicity of the equilibrium. In case of applications to both freeway and urban roads, instead, though the unicity of user equilibrium cannot be analytically proved, it is still very likely to occur.

- *User equilibrium model*

  Basing on the assumptions discussed above, a multimodal and multiclass stochastic user equilibrium assignment (SUE) model is adopted. The assignment model is formulated in a static context. A within-day dynamics version of the proposed model can be easily obtained basing on a space-discrete macroscopic or mesoscopic representation of traffic (Cascetta, 2001, 2009).

## 3.2. Multi-objective trip and road pricing models

Let $\Pi$ be the set of feasible path prices $\pi$:

$$\Pi = \left\{ \pi \in \mathbb{R}^{|K|} | \pi = (\Delta^T l) \odot \pi', \pi' \in [LB_{\pi'}, UB_{\pi'}] \subseteq \mathbb{R}^{|K|} \right\} \tag{7}$$

where $\odot$ is the element-wise product operator. According to Eq. 7, a vector of path prices is feasible if it results from the product between path lengths, i.e., $\Delta^T l$, and unit path prices $\pi'$. Unit path prices are mainly introduced for practical reasons. In fact, it can be easier to identify feasible/acceptable constraints for the range of variation of unit path prices than for the range of total path prices. Unit path prices are assumed to be constrained in the range $[LB_{\pi'}, UB_{\pi'}]$ where $LB_{\pi'} \in (-\infty, \infty)$ and $UB_{\pi'} \in (LB_{\pi'}, \infty)$. Therefore, both positive prices, i.e., tolls, and negative prices, i.e., incentives, are allowed.

The proposed trip pricing model, i.e., path-based pricing model, is formulated as the following w-GLS multi-objective optimization problem:

$$\pi^* = \arg\min_{\pi \in \Pi} \left\{ \begin{array}{c} +w_{\text{Eff}} \frac{TTS(\pi) - TTS(\mathbf{0})}{TTS(\mathbf{0})} \\ +w_{\text{Env}} \frac{TEC(\pi) - TEC(\mathbf{0})}{EC(\mathbf{0})} \\ -w_{\text{Acc}} \frac{UA(\pi) - UA(\mathbf{0})}{UA(\mathbf{0})} \\ +w_{\text{Equ(Q)}} \frac{MAPD_{\text{Q}}(\pi) - MAPD_{\text{Q}}(\mathbf{0})}{MAPD_{\text{Q}}(\mathbf{0})} \\ +w_{\text{Equ(W)}} \frac{MAPD_{\text{W}}(\pi) - MAPD_{\text{W}}(\mathbf{0})}{MAPD_{\text{W}}(\mathbf{0})} \end{array} \right\} \tag{8}$$

Subject to:

$$f^{SUE} = \sum_{\substack{q \in Q \\ m \in M \\ w \in W}} \Delta_{m,w} p^{q,m,w} \left(-\Delta_{m,w}^T c^q(f^{SUE}) - g^{nad,q,m,w} - \pi\right) \frac{\psi_w^q}{\eta^{q,m}} \cdot d_w \tag{9}$$

$$h_{\pi>0}^T \pi_{>0} - h_{\pi<0}^T |\pi_{<0}| \leq b \tag{10}$$

$$h_{\pi>0}^T \pi_{>0} \geq h_{\pi<0}^T |\pi_{<0}| \tag{11}$$

where $\pi$ is the vector of users' class-anonymous path prices, summed to the generalized path costs, and $\pi^*$ is the vector of optimal path prices.

Eq. 9 depicts the multimodal multiclass fixed-point Stochastic User Equilibrium model (see Appendix B for its derivation), which acts as a constraint.

Eq. 10, depicts a net revenue constraint where $h_{\pi \lessgtr 0}$ are the vectors of path flows corresponding to $\pi \lessgtr 0$, i.e., positive ($> 0$) and negative ($< 0$) path prices, respectively, and $b \in \mathbb{R}_+$, is the desired maximum net revenue for the pricing operator. By setting $b = 0$, a revenue-neutral scheme is obtained. Eq. 11 ensures that the total amount of collected tolls is not lower than the total amount of incentives applied.

The scalarized single-objective function in Eq. 8 is composed by five components. Each component measures the relative distance of a performance indicator between the pricing scenario and the non-pricing scenario. The five objective components relate to five different pricing goals, which are listed below. A detailed discussion of each objective, and proposed measures of performance, is provided in Sections 4.3.1 to 4.3.4:

- Traffic efficiency
- Environmental sustainability,
- Users' acceptance,
- Social equity, and
- Spatial equity.

The values of the five weights in Eq. 8, i.e., $w_{\text{Eff}}, w_{\text{Env}}, w_{\text{Acc}}, w_{\text{Equ(Q)}}, w_{\text{Equ(W)}}$ can be set by the analyst depending on the relative importance that he/she wishes to give to different pricing objectives, such that:

$$\sum_{i=\{\text{Eff},\text{Env},\text{Acc},\text{Equ(Q)},\text{Equ(W)}\}} w_i = 1 \tag{12}$$

It is worth noting that, among the five objectives, users' acceptance is the only component to be maximized in the optimal solution. For this reason, this component is summed with a negative sign in Eq. 8.

The road pricing model, i.e., link-based pricing model, is a special case of the trip pricing model in Eqs. 8-11. In particular, the road pricing model can be formulated as a trip pricing model, where the set of feasible link prices is defined by:

$$\Gamma = \left\{ \pi \in \mathbb{R}^{|K|} | \pi = \Delta^T(l \odot \gamma'), \gamma' \in \left[ LB_{\gamma'}, UB_{\gamma'} \right] \subseteq \mathbb{R}^{|A|} \right\} \tag{13}$$

Basically, Eq. 13 reveals that, in road pricing, the corresponding path prices (which are summed in Eq. 9, see the last term in the brackets) are a linear combination of link prices. In other words, from a path-based viewpoint, link prices are a source of additive path costs. Consequently, the degree of freedom of a road pricing system is always lower than the one of a trip pricing system. This evidence will affect the performances of road pricing schemes, in comparison to trip pricing ones, as shown in Section 4.

### 3.2.1. Traffic efficiency

The efficiency of a transportation system can be defined as the degree at which the users' disutility of travel is minimized. As anticipated in Section 3.2, travel disutility can be measured in terms of travel time spent in the system, travel cost, consumed energy, or combination of all these factors, i.e., in terms of user generalized cost of travel.

Two main measures of performance (MoP) have been adopted in the field literature to quantify the degree of transportation efficiency:

$$TTS(\pi) = \sum_{\substack{q \in Q \\ m \in M}} f^{q,m}(\pi)^T TT\big(f(\pi)\big) \tag{14}$$

$$TGC(\pi) = \sum_{\substack{q \in Q \\ m \in M}} f^{q,m}(\pi)^T c^q\big(f(\pi)\big) \tag{15}$$

In Eq. 14, efficiency is measured in terms of total time spent by vehicles in the system. As monetary costs are not considered in Eq. 14, the financial disutility for the users due to congestion pricing does not influence traffic efficiency. Therefore, Eq. 14 is a proxy only of the level of congestion at the network level (given convex and strictly increasing flow-travel time functions, such as the BPR).

Conversely, Eq. 15 takes into account all components of travel disutility, including travel costs, due both to consumed energy and other out-of-pocket expenses (e.g., fares, tolls). From this perspective, the total generalized cost in Eq. 15 is a proxy of social welfare, that is a measure of the net utility of travelling. Therefore, the total generalized cost may seem the most complete MoP to adopt in pricing design. In truth, this measure weights objectives with users tastes (e.g., via the VOT), thus providing a measure of users' satisfaction rather than of transportation efficiency. Notwithstanding, the degree of users' acceptance of a pricing scheme does not necessarily depends on the same attributes that influence travel choices, or with the same weights.

Therefore, as users' acceptance goal is separately considered in the multi-objective formulation of the pricing design problem (Eq. 8), in this work we adopted the total time spent in Eq. 14 as a measure of traffic efficiency.

### 3.2.2. Environmental sustainability

Environmental externalities of urban transportation are a major concern of our society (May and Milne, 2000). Congestion pricing schemes aiming at reducing transport emissions – also called environmental pricing or emission charging – has been proposed in the field literature (e.g., Johansson, 1997, 2006; Yin and Lawphongpanich, 2006; Chen and Yang, 2012; Yang et al., 2014; Cipriani et al., 2019).

Traffic efficiency and environmental efficiency may be two contrasting objectives, since the SO flow distribution based on total travel time differ from the one based on total emissions (Yin and Lawphongpanich, 2006). Therefore, an objective component that explicitly accounts for the level of emissions produced by vehicles travelling in the network is included in the multi-objective pricing design model.

In the field literature on emission charging, it is customary to adopt link-based (average) emission rate functions for each pollutant. The availability of such functions, however, is often limited, and re-estimation of pollutant-specific parameters in each case-study may be not affordable.

Therefore, in this work, we adopt the total energy consumed by vehicles travelling in the network as a proxy of the level of emissions produced in the network:

$$TEC(\pi) = \sum_{\substack{q \in Q \\ m \in M}} f^{q,m}(\pi)^T [l^T SEC^{q,m}(v)] \tag{16}$$

In Eq. 16, $SEC^{q,m}(v)$ is the specific energy consumption, i.e., the energy consumption of a vehicle class per unit of travelled distance.[2] In particular, the average speed can be computed as: $v = TT(f(\pi)) \oslash l$, where $\oslash$ is the element-wise division operator. Link-based (average) specific energy consumption functions have been calibrated in the field literature for several class of vehicles and in different operating contexts (see, for instance, Fiori et al, 2019; Fiori et al., 2021b), which improves their transferability.

### 3.2.3. Users' acceptance or perceived cost

Consensus building upon a congestion pricing scheme is critical for the success of its implementation (e.g., Viegas, 2001; Levinson, 2010; Grisolia et al. 2015). In the past fifteen years, many schemes failed because they were stopped by public referenda or citizens' petitions (e.g., Gu et al., 2018).

To address public acceptability of congestion pricing schemes, constraints to the pricing design problem accounting for users' acceptance, such as the Pareto-improving constraint, have been proposed. Such approaches however are not grounded on evidence about what users' acceptance depends on. In fact, users may be less sensitive to e.g., the reduction of travel times in the pricing scenario than to the increase of travel comfort. Eventually, users' acceptance of a congestion pricing scheme may not even be related to transportation performances, rather influenced by e.g., the simplicity of the pricing scheme (e.g., Li and Hensher, 2010), or by its social benefits (e.g., Albalate and Bel, 2009; Odeck and Kjerkreit, 2010; May et al., 2010).

---

[2] Please note that a vehicle class is identified by the type of vehicle selected by users of class *q* to travel on mode *m*. In this view, users' classes can be seen as a combination of socio-economic variables and vehicle consumption parameters.

Though strategies to increase public acceptance are available and have been tested, quantitative approaches to measure public acceptance are not well established in the literature (available studies focused on ex-post evaluations only, rather than on ex-ante interviews to support pricing design).

In this work, we proposed two methodological approaches to bridge this gap:

- a bottom-up approach, which consists in the specification of users' acceptance function based on evidence collected through Stated Preferences surveys, and estimation of its parameters;
- a top-down approach, which consists in the formulation of users' acceptance function based on consolidated theories to interpret users' behaviours (such as RU models).

The former approach is original to the field, and it is authors' opinion that it is the elective approach to be pursued in future research. In turn, it requires case-dependent analyses (see, e.g., Milenković et al., 2019). On the other hand, behavioural theories, which explain users' choices in the usage of the transportation supply, are well-consolidated and provides measures to quantify the level of users' satisfaction, given a set of path/mode alternatives.

In this work, we incorporated a measure based on RU theory in the multi-objective pricing design model. Specifically, for users' class $q$, the expected value of the maximum perceived utility, $s_w^q(V^{K_w|q})$, i.e., satisfaction function of path costs (over different modes) between OD pair $w$ is adopted as a measure of (average) generalized costs as in e.g., Zhao and Kockelman (2006), Wu et al. (2011), Liu et al. (2014) :

$$s_w^q(V^{K_w|q}) = E\left\{\max_{k \in K_w} U_k^{K_w|q}\right\} \tag{17}$$

where $U_k^{K_w|q}$ is the perceived utility, for users of class $q$, of a path alternative $k$ between OD pair $w$.

Depending on the path choice model, the satisfaction function can have a closed-form expression or not (e.g., in case of a multinomial logit model, it coincides with the *logsum* function; Ben-Akiva and Lerman, 1975).

Let $s_w$ be the total satisfaction across all users' classes between OD pair $w$:

$$s_w = \sum_{q \in Q} s_w^q \tag{18}$$

The adopted users' acceptance measure is given by:

$$UA = \sum_{w \in W} s_w \tag{19}$$

Obviously, the model could easily incorporate other users' acceptance functions derived from preference studies.

### 3.2.4. Social and spatial equity

Heterogeneity in travel cost for the same mode/route alternative among different users' classes, due for instance to VOT variability, is a clear symptom of the inherent social inequity of the transportation system. Once congestion pricing schemes are applied, the gap between user classes inevitably widens, making the system even more inequitable. In addition, users moving between certain OD pairs may not have mode/path alternatives as competitive as those available between other OD pairs. *Spatial inequality* is *structurally inherent to the system*, whether pricing is applied or not. In fact, it depends on the availability of path/mode alternatives connecting OD pairs, which structurally depends on the topology of the network and on the transportation system. Needless to say, isotropic path prices, i.e., not differentiated by OD pairs – as in road pricing schemes – exacerbates spatial inequality effects.

To explicitly consider these effects in the pricing design problem, measures aiming at quantifying the degree of social and spatial inequity have been proposed (e.g., Yang and Zang, 2002; Yin and Yang, 2004). In particular, in the majority of the approaches, constraints based on these measures are introduced to limit the amplification of the inequity effects relative to the non-pricing scenario. As discussed in Section 2, a constrained optimization, however, does not optimize equity simultaneously with all other pricing objectives.

In contrast, a multi-objective design problem can effectively consider equity and other objectives at the same time. In this perspective, the designed tolls and incentives may even improve the social and spatial equity of the transportation system, compared to the pre-existing scenario.

Among proposed measures of equity, the Gini coefficient has been by far the most adopted in the field literature (van Wee and Mouter, 2021). Variations of the Gini coefficient, such as the Theil index, the Palma ratio, the Atkinson index and the Suit index have also been specified (for a review of applications see Souche et al., 2016). These indicators, however, have been mostly applied to measure social inequality effects. Rather, accessibility measures have been mostly applied to quantify spatial inequalities (e.g., Souche et al., 2016; Zhong et al., 2021).

In this work we present a novel equity measure, which can be specialized to capture both spatial and social effects. The proposed measure captures the average dispersion of the user satisfaction (about available transportation alternatives), across OD pairs and across users' classes. The adoption of user satisfaction as quantity of interest in the evaluation of equity is novel to the field.

Assuming that users' perceived cost is proportionally higher on long-distance OD pairs than on short-distance ones, the satisfaction $s_w^q$ in Eq. 17 is normalized by trip length in order to compare satisfaction of users of class $q$ among different ODs:

$$s_w^{q\,'} = \frac{|K_w|}{\sum_{k \in K_w} \Delta_{\cdot,k}^T l} s_w^q \tag{20}$$

$$s_w' = \frac{1}{|Q|} \sum_{q \in Q} s_w^{q\,'} \tag{21}$$

$$s^{q\,'} = \frac{1}{|W|} \sum_{w \in W} s_w^{q\,'} \tag{22}$$

The spatial equity is thus measured by the mean percentage deviation, over OD pairs, of the unit satisfaction in Eq. 21:

$$MAPD_W = \frac{1}{|W|} \sum_{w \in W} \left| \frac{s_w' - \overline{s_W'}}{\overline{s_W'}} \right| \tag{23}$$

where $\overline{s_W'}$ is the average unit satisfaction among all OD pairs, i.e., $\overline{s_W'} = \frac{1}{|W|} \sum_{w \in W} s_w'$. Similarly, the social equity is measured by the mean percentage deviation, over users' classes, of the unit satisfaction in Eq. 22:

$$MAPD_Q = \frac{1}{|Q|} \sum_{q \in Q} \left| \frac{s^{q\,'} - \overline{s^{Q'}}}{\overline{s^{Q'}}} \right| \tag{24}$$

where $\overline{s^{Q'}}$ is the average unit satisfaction among users' classes, i.e., $\overline{s^{Q'}} = \frac{1}{|Q|} \sum_{q \in Q} s^{q\,'}$.

The mean percentage deviation measure differs from the Gini Coefficient, as the former computes deviations of each OD pair (or user class), from the center of mass, while the latter computes the mutual deviations from one element to all the others. It is worth noting that the proposed measure enhances and generalizes the one adopted in Zheng and Geroliminis (2020) to evaluate spatial equity.

Eventually, social and spatial equity measures in Eqs. 23 and 24, can be combined in a single measure to quantify equity effects over both dimensions, in order to reduce its weight in the multi-objective optimization problem:

$$MAPD = \frac{1}{|W||Q|} \sum_{\substack{q \in W \\ w \in W}} \left| \frac{s_w^{q\,'} - \overline{s_W^{q\,'}}}{s_W^{q\,'}} \right| \tag{25}$$

where $\overline{s_W^{q\,\prime}} = \frac{1}{|Q|} \sum_{q \in Q} s_w^{q\,\prime}$.

## 4. NUMERICAL EXPERIMENTS

In this section, a thorough comparison of performances of road (link-based) and trip (path-based) pricing models is performed on the Nguyen-Dupuis network (Nguyen and Dupuis, 1984). Models have been compared both in a car-only network with fixed demand, and in a multimodal network with mode choice elastic demand. Pricing schemes were evaluated under a multiclass stochastic traffic assignment at equilibrium.

In order to quantify the impact of design criteria on pricing performances, the comparison analysis included models designed either with a single objective (traffic efficiency, environment, users' acceptance, social equity, spatial equity) or with multiple (all) objectives. Eventually, a comparison between pricing schemes and their revenue-neutral variations was carried out.

It is worth clarifying that the objective of this numerical exercise was purely comparative. We aimed at *i)* quantifying the impact of different model variants on the pricing scheme performances and *ii)* identifying the range of variability of performances among different design configurations. It comes unsaid that the absolute value of models' performances measured in this experiment are not intended to be generally transferable to different network configurations.

### 4.1. Network and models

The multimodal Nguyen-Dupuis network used in this work is depicted in Figure 1. The original road network was extended in order to include two alternative modes, e-bikes on dedicated lanes and a metro line. Four OD pairs were considered (AD, BD, AC, BC), each served by a different number of alternative modes, so that the spatial equity effects can be easily outlined.

The road network is composed by 2-lane tolled highway, 2-lane secondary urban roads with dedicated bike lanes, and 1-lane local urban roads without bike lanes. Road paths connects all four OD pairs and are used by Internal Combustion Engine (ICE) vehicles, while the bike lanes connect only two OD pairs (AD and AC) and are used by e-Bikes. The metro line serves two OD pairs (AD and BD). Paths are listed in Figure 1.

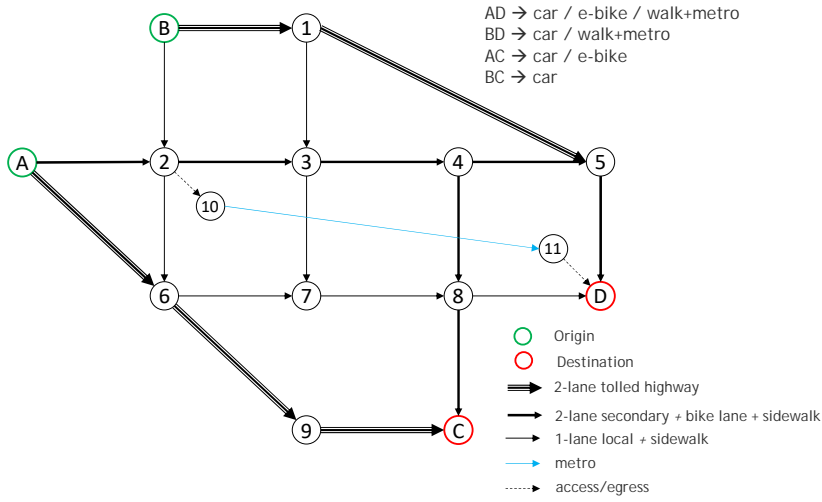

AD → car / e-bike / walk+metro
BD → car / walk+metro
AC → car / e-bike
BC → car

| LINK | length [km] | capacity [veh/h] | CAR speed [km/h] | e-BIKE speed [km/h] | WALK+METRO speed [km/h] |
|---|---|---|---|---|---|
| B,1 | 3 | 3600 | 120 | - | - |
| 1,5 | 5 | 3600 | 120 | - | - |
| A,6 | 3 | 3600 | 120 | - | - |
| 6,9 | 3 | 3600 | 120 | - | - |
| 9,C | 1 | 3600 | 120 | - | - |
| A,2 | 1 | 2400 | 50 | 15 | 5 |
| 2,3 | 1 | 2400 | 50 | 15 | 5 |
| 3,4 | 1 | 2400 | 50 | 15 | 5 |
| 4,5 | 1 | 2400 | 50 | 15 | 5 |
| 5,D | 1 | 2400 | 50 | 15 | 5 |
| 4,8 | 1 | 2400 | 50 | 15 | 5 |
| 8,C | 1 | 2400 | 50 | 15 | 5 |
| B,2 | 1 | 1600 | 30 | - | 5 |
| 1,3 | 1 | 1600 | 30 | - | 5 |
| 2,6 | 1 | 1600 | 30 | - | 5 |
| 3,7 | 1 | 1600 | 30 | - | 5 |
| 6,7 | 1 | 1600 | 30 | - | 5 |
| 7,8 | 1 | 1600 | 30 | - | 5 |
| 8,D | 1 | 1600 | 30 | - | 5 |
| 2,10 | 0.3 | - | - | - | 5 |
| 10,11 | 4 | - | - | - | 70 |
| 11,D | 0.3 | - | - | - | 5 |

○ Origin
○ Destination
■ 2-lane tolled highway
→ 2-lane secondary + bike lane + sidewalk
→ 1-lane local + sidewalk
→ metro
⋯ access/egress

| ID | OD | MODE | Nodes |
|---|---|---|---|
| 1 | AD | Car | A,2,3,4,5,D |
| 2 | AD | Car | A,2,3,4,8,D |
| 3 | AD | Car | A,2,3,7,8,D |
| 4 | AD | Car | A,2,6,7,8,D |
| 5 | AD | Car (Freeway) | A,6,7,8,D |
| 26 | AD | e-Bike | A,2,3,4,5,D |
| 28 | AD | Metro | A,2,10,11,D |

| ID | OD | MODE | Nodes |
|---|---|---|---|
| 6 | BD | Car (Freeway) | B,1,5,D |
| 7 | BD | Car (Freeway) | B,1,3,4,5,D |
| 8 | BD | Car | B,1,3,4,8,D |
| 9 | BD | Car (Freeway) | B,1,3,7,8,D |
| 10 | BD | Car | B,2,3,4,5,D |
| 11 | BD | Car | B,2,3,4,8,D |
| 12 | BD | Car | B,2,3,7,8,D |
| 13 | BD | Car | B,2,6,7,8,D |
| 29 | BD | Metro | B,2,10,11,D |

| ID | OD | MODE | Nodes |
|---|---|---|---|
| 14 | AC | Car (Freeway) | A,6,9,C |
| 15 | AC | Car (Freeway) | A,6,7,8,C |
| 16 | AC | Car | A,2,3,4,8,C |
| 17 | AC | Car | A,2,3,7,8,C |
| 18 | AC | Car | A,2,6,7,8,C |
| 19 | AC | Car (Freeway) | A,2,6,9,C |
| 27 | AC | e-Bike | A,2,3,4,8,C |

| ID | OD | MODE | Nodes |
|---|---|---|---|
| 20 | BC | Car (Freeway) | B,1,3,4,8,C |
| 21 | BC | Car (Freeway) | B,1,3,7,8,C |
| 22 | BC | Car | B,2,3,4,8,C |
| 23 | BC | Car | B,2,3,7,8,C |
| 24 | BC | Car | B,2,6,7,8,C |
| 25 | BC | Car (Freeway) | B,2,6,9,C |

**Figure 1 – Multimodal Nguyen-Dupuis network.**

Concerning the demand model, users' mode and route choices are modelled through a single-level hierarchical logit model. At the path choice level, a c-logit model is assumed for car paths in order to account for users' perception of overlapping paths (Cascetta et al., 1996).

The utility function of a mode $m$ alternative is specified as the EMPU among path alternatives, which is known as *logsum* variable of path choices in case of a logit-based path choice model (Ben-Akiva and Lerman, 1975):

$$V_m^{M|q,w} = \frac{1}{\theta^K} \ln\left(\sum_{k \in K_{m,w}} e^{\frac{V_k^{K_{m,w}|q}}{\theta^K}}\right), \ \forall m \in M : |K_{m,w}| > 0 \qquad (26)$$

where $V_k^{K_{m,w}|q}$ is given by Eq. 27:

$$V_k^{K_{m,w}|q} = -g_k^{q,m,w} - \beta_{SF} \ln\left(SF_k^{q,m,w}\right) \qquad (27)$$

In Eq. 27, $SF_k^{q,m,w}$ is the so-called "commonality factor" of the c-logit model (Cascetta et al., 1996), i.e., a factor proportional to the overlapping degree of path $k$ with all the other available paths connecting the same OD pair, i.e., $K_{m,w}$:

$$SF_k^{q,m,w} = \sum_{j \in K_{m,w}} \left(\frac{g_{kj}^{q,m,w}}{\sqrt{g_k^{q,m,w} g_j^{q,m,w}}}\right)^{\alpha_{SF}} \qquad (28)$$

where $g_{kj}^{q,m,w}$ is equal to the sum of generalized costs perceived by users of class $q$ related to the subset of path links shared by paths $j$ and path $k$. For e-Bike and metro paths, $SF$ is assumed to be zero.

The choice probability for users of class $q$ of mode $m$ between OD pair $w$ is thus equal to:

$$p_m^{\text{M}|q,w} = e^{\frac{v_m^{\text{M}|q,w}}{\theta^{\text{M}}}} \bigg/ \sum_{j \in \text{M} : |\text{K}_{m,w}| > 0} e^{\frac{v_j^{\text{M}|q,w}}{\theta^{\text{M}}}}, \forall m \in \text{M} : |\text{K}_{m,w}| > 0 \tag{29}$$

The choice probability for users of class $q$ of path $k$ between OD pair $w$, conditional to the choice of mode $m$, is equal to:

$$p_k^{\text{K}_{m,w}|q} = e^{\frac{v_k^{\text{K}_{m,w}|q}}{\theta^{\text{K}}}} \bigg/ \sum_{j \in \text{K}_{m,w}} e^{\frac{v_j^{\text{K}_{m,w}|q}}{\theta^{\text{K}}}}, \forall k \in \text{K}_{m,w} \tag{30}$$

Concerning the supply model, the link generalized cost function $c^q(f_a)$ includes the BPR function which accounts for travel time variation with flow, a constant waiting time function and a monetary cost. For a link $a$, it is given by a linear combination of travel times and travel costs:

$$c^q(f_a) = \beta_{TT,a}[VOT^q \cdot TT(f_a) + VOWT^q \cdot WT(f_a)] + MC^{q,m}(f_a) \tag{31}$$

where $\beta_{TT,a}$ is the reciprocal substitution coefficient of travel time, $TT(f_a) = \frac{l_a}{v_a}\left[1 + \alpha_{BPR,a}\left(\frac{f_a}{cap_a}\right)^{\beta_{BPR,a}}\right]$ is the BPR function for link $a$ with parameters $\alpha_a$ and $\beta_a$ that depends on the link type/mode (e.g., for motorway $\alpha_a = 0.5$ and $\beta_a = 4$, for urban roads $\alpha_a = 2$ and $\beta_a = 4$; please note that transit and bicycle links are assumed to be uncongested, i.e., $TT(f_a) = \frac{l_a}{v_a}$ which means $\alpha_a = 0$).

$WT(f_a)$ is the waiting time function. For the sake of simplicity, the waiting time on transit paths is assumed to be independent from the transit line frequency, and thus are here assumed to be constant.

The monetary cost function of link $a$ is given by $MC^{q,m}(f_a) = \zeta^{q,m} \cdot SEC^{q,m}(l_a, f_a) \cdot l_a$ is, which converts the energy consumption of link $a$ in monetary cost via $\zeta^{qm}$ which depends on the link type/mode (road, transit, bicycle) and on the user/vehicle class (e.g., electric vehicles, internal combustion engine vehicles). The specific energy consumption function $SEC^{q,m}(l_a, f_a)$ depends on the link type/mode and user/vehicle class. The following specific energy consumption functions for each mode were adopted:

$$SEC_{ice}(v) = 0.136 + 7.04 \cdot 10^{-6} \cdot v^2 - 1.42 \cdot 10^{-3} \cdot v \tag{32}$$

$$SEC_{e-bike} = 0.10 \tag{33}$$

In Eq. 32, the average link speed $v$ is expressed in km/h and the resulting specific energy consumption is expressed in liters/veh-km (Fiori et al., 2019; values in kWh/veh-km are obtained multiplying consumptions by 8.9 kWh/l). In Eq. 33, instead, the specific energy consumption is expressed in kWh/pax-km.

Please note that the monetary cost for the users due to the consumed energy by trains operating in the metro system does not influence users' mode choice, i.e., it equals zero. Notwithstanding, a specific energy consumption function for metro trains was required in order to compute the energy consumed to serve the passenger demand on the metro line, in case of pricing schemes designed with the objective of environmental sustainability:

$$SEC_{metro} = 0.08 \tag{34}$$

where the specific energy consumption in Eq. 34 is returned in kWh/pax-km.

Two users' classes were considered in the numerical experiments, with different values of travel/waiting time.

The full list of parameter values applied in numerical experiments is provided in Table 3.

| Parameter | Unit of measurement | Value |
|---|---|---|
| $\beta_{TT}^{car}$ | | 1.0 |
| $\beta_{TT}^{e-bike}$ | | 3.0 |
| $\beta_{TT}^{metro}$ | | 1.5 |
| $\alpha_{BPR}^{highway}$ | | 0.15 |
| $\alpha_{BPR}^{urban}$ | | 2.00 |
| $\beta_{BPR}$ | | 4 |
| $\beta_{SF}$ | | 1 |
| $\alpha_{SF}$ | | 1 |
| $\zeta^{fuel}$ | €/l | 1.60 |
| $\zeta^{elect.}$ | €/kWh | 0.25 |
| $\eta^{car}$ | pax/veh | 1.20 |
| $fare^{car}$ | €/km | 0.08 |
| $fare^{metro}$ | €/pax | 2.00 |
| $VOT^{class\,1}$ | €/h | 5 |
| $VOT^{class\,2}$ | €/h | 10 |
| $VOWT^{class\,1}$ | €/h | 10 |
| $VOWT^{class\,2}$ | €/h | 20 |
| $\psi^{class\,1}$ | | 0.7 |
| $\psi^{class\,2}$ | | 0.3 |
| $WT^{car}$ | h/pax | 0 |
| $WT^{e-bike}$ | h/pax | 0 |
| $WT^{metro}$ | h/pax | 0.067 |
| $\theta^{K}$ | | 5 |
| $\theta^{M}$ | | 1 |

**Table 3 – Values of model parameters.**

## 4.2. Design of Experiments

Road and trip pricing schemes, as well as their revenue-neutral variants, were compared in two network scenarios: *i)* a *car-only network* scenario composed by car links only, with a fixed demand, and *ii)* a *multimodal network* scenario with elastic demand on mode choice.

In order to compare results of the same model in different network scenarios, demand values assigned in the multimodal network have been set equal to values that return SUE car flows (without pricing) equal to the car demand flow applied in the car-only network scenario.

In addition, in the *multimodal network* scenario, road and trip pricing schemes were compared under both first-best and second-best conditions. In first-best conditions, all network links/paths were considered priceable, including e-bike and metro elements, while in second-best conditions, pricing was restricted only to car network elements. In this network scenario, the first-best pricing scheme will also be referred to as *multimodal pricing* scheme, while the second-best pricing scheme will also be referred to as *monomodal pricing* scheme.

All pricing schemes were designed either considering a single-objective or considering all objectives simultaneously (in this case, equal weights in Eq. 8 were adopted). The black-box optimization problem in Eqs. 8-11 was solved through both the genetic algorithm and the interior-point method coded in Matlab®, in order to assess the robustness of the optimal solution. In particular, both the algorithms converged to the same optimal solution in 92% of cases (in the remaining ones, we selected the best solution between the two, i.e., the solution with the lowest objective function value).

Pricing model performances were evaluated using indicators discussed in Section 3.2, i.e., the total travel time spent in the network (TTS), the total energy consumed for travelling by car, e-Bike and metro (TEC), the user's perceived cost of travelling (PC, that is the inverse of UA), the mean absolute percentage deviation of generalized costs across users' classes (MAPD$_Q$) and across OD pairs (MAPD$_W$).

In addition, performances were also evaluated in terms of total network traffic per mode, flow-capacity ratio in the car network, percentage of modal split of e-Bike and metro modes, and revenues for the pricing operator, for the highway operator and for the metro system operator.

As discussed in Section 3.2, model performances were calculated in terms of relative distance between the scenarios with pricing and the scenario without pricing – in the following referred to as reference scenario.

Results of pricing scheme performances in the two mentioned scenarios are discussed in Section 4.3 and Section 4.4, respectively.

### 4.3. Results in the car-only network

*4.3.1. Reference scenario*

Figure 2 depicts the SUE link flow distribution in the scenario without pricing. The overall level of congestion is high, with most of the link with a flow-capacity ratio greater than 75%. The total travel time spent in the system is 3.51 h (26 min/pax, on average).

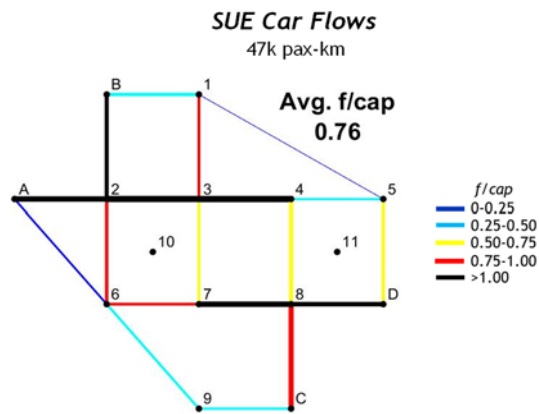

**Figure 2 – SUE flow distribution and passenger traffic in the reference scenario, i.e., (car-only, i.e., monomodal) multiclass SUE without pricing. Network links are coloured according to the flow/capacity ratio. Total OD flow on each OD pair is equal to 2,000 pax/h (network flow equal to 8,000 pax/h).**

*4.3.2. Road vs. trip pricing schemes*

Road and trip pricing schemes were designed via Eqs. 8-9, either with a single objective or with multiple objectives. Link/path unit prices were constrained in the range [0,5] €/km.[3]

A first comparison between the pricing schemes was carried out basing on the optimal prices resulting from traffic efficiency optimization, i.e., $w_{\mathrm{Eff}} = 1$ and remaining weights set to zero in Eq. 8. Table 4 reports the SUE flow distribution in the reference scenario (without pricing), and in the scenarios with optimal road and trip pricing.

Results revealed that, in road pricing, link price values yielding a high total path price do not imply low path flows. In trip pricing, instead, high path prices always return low path flows. Consequently, path flows are also more dispersed in road pricing than in trip pricing, no matter how much high the prices are.

---

[3] Theoretically, a constraint on the price space may prevent the existence of a feasible price solution that leads to the SO condition (and corresponding flow distribution). This hypothesis was verified experimentally in terms of TTS, and results showed that the error between the optimal TTS resulting from trip pricing schemes and the theoretical SO (i.e., without price constraints) was less than 3%. Concerning road pricing, instead, the optimal TTS was always found greater than the optimal TTS in trip pricing, due to the effect of link price on path flows (see the paper text for an explanation).

| OD | Path | Reference | | Road pricing (efficiency) | | | Trip pricing (efficiency) | | |
|---|---|---|---|---|---|---|---|---|---|
| | | *Flow [pax/h]* | *TT [min]* | *Flow [pax/h]* | *TT [min]* | *Price [€/km]* | *Flow [pax/h]* | *TT [min]* | *Price [€/km]* |
| AD | 1 | 598 | 17 | 646 | 8 | 3.3 | 572 | 8 | 1.5 |
| | 2 | 462 | 28 | 377 | 10 | 4.4 | 501 | 9 | 1.7 |
| | 3 | 303 | 48 | 108 | 11 | 5.0 | 17 | 10 | 5.0 |
| | 4 | 300 | 48 | 102 | 14 | 5.0 | 17 | 12 | 5.0 |
| | 5 | 337 | 41 | 767 | 10 | 2.5 | 893 | 9 | 0.8 |
| BD | 6 | 374 | 5 | 1553 | 7 | 0.0 | 1935 | 7 | 0.0 |
| | 7 | 341 | 13 | 186 | 10 | 1.5 | 2 | 9 | 4.9 |
| | 8 | 264 | 24 | 108 | 11 | 1.9 | 2 | 10 | 4.9 |
| | 9 | 174 | 44 | 31 | 12 | 2.8 | 2 | 11 | 4.9 |
| | 10 | 302 | 22 | 64 | 10 | 4.3 | 15 | 9 | 5.0 |
| | 11 | 235 | 33 | 37 | 12 | 4.8 | 15 | 10 | 5.0 |
| | 12 | 156 | 52 | 11 | 13 | 5.0 | 16 | 11 | 4.9 |
| | 13 | 155 | 53 | 10 | 15 | 5.0 | 14 | 13 | 5.0 |
| AC | 14 | 476 | 4 | 1642 | 4 | 0.0 | 1948 | 4 | 0.0 |
| | 15 | 266 | 32 | 79 | 8 | 2.5 | 5 | 7 | 4.9 |
| | 16 | 362 | 20 | 39 | 7 | 4.8 | 16 | 7 | 4.9 |
| | 17 | 239 | 39 | 11 | 9 | 5.0 | 13 | 8 | 5.0 |
| | 18 | 237 | 39 | 10 | 11 | 5.0 | 13 | 10 | 5.0 |
| | 19 | 420 | 11 | 218 | 7 | 1.7 | 5 | 6 | 5.0 |
| BC | 20 | 423 | 15 | 531 | 8 | 1.9 | 856 | 8 | 0.3 |
| | 21 | 276 | 35 | 152 | 10 | 2.8 | 45 | 9 | 2.4 |
| | 22 | 375 | 24 | 183 | 9 | 3.8 | 9 | 8 | 5.0 |
| | 23 | 247 | 43 | 52 | 11 | 5.0 | 10 | 9 | 4.9 |
| | 24 | 245 | 44 | 50 | 13 | 5.0 | 9 | 11 | 5.0 |
| | 25 | 435 | 15 | 1032 | 9 | 1.7 | 1070 | 7 | 0.2 |

**Table 4 – Road vs. trip pricing schemes, designed with traffic efficiency as single objective. Flow values refer to the sum of values among users' classes. Red boxes highlight paths which are priced more than 4/km. Grey text highlight paths which are priced more than 4€/km or are assigned with lower than 20 pax/h, at SUE condition with optimal pricing.**

Performance comparison of road and trip pricing schemes designed with all tested objective functions is reported in Table 5. Concerning the measure of performance discussed in Section 3.2, it is clear that the minimum value of the relative error on each measure is obtained by designing the pricing scheme with that measure as single objective. For instance, the minimum value of $\Delta$TTS is obtained when the pricing scheme is designed aiming at traffic efficiency only, i.e., $w_{\text{Eff}} = 1$ and remaining weights set to zero.

| MoP | Ref. | Road pricing | | | | | | Trip pricing | | | | | |
|---|---|---|---|---|---|---|---|---|---|---|---|---|---|
| | | *Objective function* | | | | | | *Objective function* | | | | | |
| | | *Eff.* | *Env.* | *Acpt.* | *Soc. Equ.* | *Spat. Equ.* | *All* | *Eff.* | *Env.* | *Acpt.* | *Soc. Equ.* | *Spat. Equ.* | *All* |
| ΔTTS | | -71% | -19% | - | -46% | +96% | -40% | -74% | 0% | -62% | -39% | +136% | -73% |
| ΔTEC | | -9% | -10% | - | -4% | +4% | -4% | -9% | -12% | -6% | -5% | +9% | -9% |
| ΔPC | | +78% | +118% | - | +302% | +210% | +68% | +14% | +21% | -4% | +288% | +128% | +13% |
| ΔMAPD_Q | | -90% | -80% | - | -95% | -66% | -81% | -77% | -28% | -12% | -96% | -38% | -77% |
| ΔMAPD_W | | +132% | +37% | - | -47% | -89% | -68% | +70% | +37% | -24% | -41% | -100% | -100% |
| *Average* | | *+8%* | *+9%* | - | *+22%* | *+31%* | *-25%* | *-15%* | *4%* | *-12%* | *+22%* | *+27%* | *-49%* |
| Avg. Travel Time [min/pax] | 26 | 8 (-71%) | 21 (-19%) | - | 14 (-46%) | 51 (+96%) | 16 (-40%) | 7 (-74%) | 26 (0%) | 10 (-62%) | 16 (-39%) | 61 (+136%) | 7 (-73%) |
| Traffic [kpax-km] | 47 | 54 (+15%) | 49 (+4%) | - | 50 (+6%) | 43 (-8%) | 48 (+2%) | 55 (+17%) | 49 (+4%) | 52 (+10%) | 51 (+9%) | 45 (-4%) | 54 (+14%) |
| Avg. f/cap ratio | 0.76 | 0.57 (-25%) | 0.62 (-18%) | - | 0.69 (-9%) | 0.82 (+8%) | 0.70 (-8%) | 0.53 (-30%) | 0.60 (-21%) | 0.64 (-16%) | 0.68 (-11%) | 0.82 (+8%) | 0.56 (-26%) |
| Total revenue from pricing [k€] | - | 72 | 85 | - | 171 | 108 | 77 | 21 | 10 | 2.0 | 166 | 61 | 29 |
| Total revenue from highway tolls [k€] | 1.3 | 2.8 (+115%) | 2.3 (+76%) | - | 1.9 (+46%) | 0.6 (-54%) | 1.6 (+23%) | 3.1 (+138%) | 2.4 (+85%) | 2.2 (+69%) | 2.0 (+54%) | 0.8 (-38%) | 2.9 (+123%) |
| Revenue per pax from pricing [€/pax] | - | 8.9 | 10.7 | - | 21.4 | 13.5 | 9.5 | 2.7 | 1.2 | 0.3 | 20.8 | 7.6 | 3.6 |

*Eff = traffic efficiency (min TTS), Env = environment (min TEC), Acpt. = users' acceptance (max UA), Soc. Equ = social equity (min MAPD_Q), Spat. Equ = spatial equity (min MAPD_W), All = all objectives (with equal weights).*

**Table 5 – Performance comparison of road and trip pricing schemes, designed with either a single objective or with all objectives. In top half of the table, performances are measured through the cost measures discussed in Section 3.2: total travel time (TTS), total energy consumption (TEC), perceived cost (PC, i.e., the inverse of UA), mean absolute percentage deviation of generalized costs across users' classes (MAPD_Q) and OD pairs (MAPD_W). In the bottom half of the table, performances of the pricing schemes are evaluated in terms of total network traffic, and of revenues for the pricing and the highway operators. Variations of performance are computed from the reference scenario, i.e., SUE condition without pricing. Improvements and deteriorations of a measure of performance are indicated with green and red text, respectively.**

From the table, the following main considerations are worth noting:

- Performances of road pricing are consistently lower than trip pricing ones. In fact, trip pricing is more efficient, environmentally sustainable, acceptable, and equitable than road pricing. In addition, at the path-level, the total path price paid by users in road pricing is always greater than the one paid in trip pricing. In fact, the total revenue for the pricing operator, as well as the average price paid by the user, is always greater in road pricing than in trip pricing. This result is in accordance with the numerical evidence provided in Zangui et al. (2015a). Such behaviour is basically due to the fact that the link price feasibility space is included in the path price feasibility space, and thus the degree of freedom of a road pricing system is lower than the one of a trip pricing system.

- User's acceptance is the most challenging MoP to optimize. This is due to the fact that the pricing model aiming at all other objectives increases the cost paid by travellers by a quantity which is higher than the corresponding travel time savings. In particular, there is no price configuration in road pricing schemes which allows increasing users' satisfaction about travel alternatives, not even when the scheme is designed with users' acceptance as single objective. In turn, trip pricing schemes can lead only to marginal improvements with respect to the reference scenario. From this perspective, revenue-neutral schemes, that allow establishing either tolls or incentives on network elements, are thus expected to be more effective than pricing schemes in increasing users' satisfaction.

- It is harder to increase spatial equity than social equity. In fact, while social equity can be controlled by pricing, e.g., setting a high price on all links is sufficient to increase social equity (at the limit, if all prices were infinite, all users would experience the same cost), spatial equity structurally depends on the topology of the network rather than on transportation system performances (and thus it is harder to control with pricing). Spatial equity and traffic efficiency revealed to be very contrasting objectives, i.e., a decrease of total travel time is detrimental to spatial equity, and vice versa (the highest TTS reduction, -71% and -74% in road and trip pricing, respectively, corresponds to the highest increase of MAPDw, +132% and +70%). In fact, if pricing scheme aims at maximizing spatial equity alone, the shift of users with alternative paths from shortest to longer/less attractive paths in order to freeing capacity for those users without alternatives, deteriorates the total time spent in the system, as well as other MoPs than equity. This evidence is a further confirmation of the need to adopt a multi-objective framework to design pricing schemes.

- If a multi-objective framework is adopted to design pricing schemes, i.e., all MoPs are included in the objective function, the average performance of the system (either with road or trip pricing) improves on all objectives, with respect to single-objective optimization. In fact, the maximum average decrease on all objectives is achieved in a multi-objective optimization, reaching -25% and -49% for road and trip pricing, respectively. However, in road pricing, the optimal solution is very influenced by the users' acceptance performance (a limited deterioration of users' acceptance relative to the reference scenario, i.e., +68% in the table, is paid by halving the improvement on total travel time with respect to the maximum improvement, i.e., -40% in multi-objective optimization vs. -71% in traffic efficiency optimization). This effect is not observed in trip pricing, where the TTS value in multi-objective optimization is very closed to the optimal value reached in traffic efficiency optimization, and the corresponding network traffic reaches the same value. Furthermore, the increase of revenues (relative to the reference scenario) for the highway operator is lower in multi-objective optimization than in traffic efficiency optimization. This result revealed that traffic is moved from highways to urban roads in order to re-balance (spatial) equity/users' satisfaction and traffic efficiency (see also the previous comment on contrasting objectives).

- Improvements on energy consumptions are less relevant than on other MoPs. In fact, being the demand fixed, i.e., the demand is assigned to a car-only network, the number of circulating cars is constant, and the limited increase in pax-km, i.e., the increase of traffic on longer paths, is less beneficial on energy consumption, than on other MoPs (such as spatial equity). This finding reveals that pricing schemes may have limited impact in car-only networks, while stronger impacts are expected in multimodal networks where more energy-efficient mode alternatives are available.

### 4.3.3. Pricing vs. Revenue-neutral schemes

By relaxing the non-negativity constraint on the lower bound of unit prices and adding nonlinear constraints in Eqs. 10-11 to the design problem, revenue-neutral schemes were obtained. Accordingly, link/path unit prices were constrained in the range [−5,5] €/km, and the balance constant in Eq. 10, which corresponds to the maximum net revenue allowed for the pricing operator, was assumed equal to 1000 € (a zero value for net revenues would have unnecessarily increased the computational burden).

Results from pricing and revenue-neutral schemes design are reported in Table 6. In terms of pricing objectives, the comparative analysis was carried out between single-objective optimization on traffic efficiency and multi-objective optimization. For the sake of comparison, corresponding columns for road and trip pricing schemes were taken from Table 5.

| MoP | Ref. | Road pricing *Obj. function* | | Trip pricing *Obj. function* | | Revenue-neutral Road pricing *Obj. function* | | Revenue-neutral Trip pricing *Obj. function* | |
|---|---|---|---|---|---|---|---|---|---|
| | | *Eff.* | *All* | *Eff.* | *All* | *Eff.* | *All* | *Eff.* | *All* |
| ΔTTS | | -71% | -40% | -74% | -73% | -46% | +19% | -68% | -6% |
| ΔTEC | | -9% | -4% | -9% | -9% | -4% | +1% | -7% | -1% |
| ΔPC | | +78% | +68% | +14% | +13% | -11% | +6% | -17% | +7% |
| $\Delta MAPD_Q$ | | -90% | -81% | -77% | -77% | +1300% | -6% | +13% | -5% |
| $\Delta MAPD_W$ | | +132% | -68% | +70% | -100% | +740% | -65% | +102% | -88% |
| *Average* | | *+8%* | *-25%* | *-15%* | *-49%* | *+47%* | *-9%* | *+5%* | *-19%* |
| Avg. Travel Time [min/pax] | 26 | 8 (-71%) | 16 (-40%) | 7 (-74%) | 7 (-73%) | 14 (-46%) | 31 (+19%) | 8 (-68%) | 24 (-6%) |
| Traffic [kpax-km] | 47 | 54 (+15%) | 48 (+2%) | 55 (+17%) | 54 (+14%) | 50 (+6%) | 46 (-2%) | 53 (+13%) | 48 (+2%) |
| Avg. f/cap ratio | 0.76 | 0.57 (-25%) | 0.70 (-8%) | 0.50 (-30%) | 0.56 (-26%) | 0.68 (-11%) | 0.78 (+3%) | 0.61 (-20%) | 0.75 (-1%) |
| Total net revenue from pricing [k€] | - | 72 | 77 | 21 | 29 | 0.9 | 1.0 | 0.0 | 0.9 |
| Total revenue from highway tolls [k€] | 1.3 | 2.8 (+115%) | 1.6 (+23%) | 3.1 (+138%) | 2.9 (+123%) | 1.9 (+46%) | 1.1 (-15%) | 2.5 (+92%) | 1.3 (+2%) |
| Net revenue per pax from pricing [€/pax] | - | 8.9 | 9.5 | 2.7 | 3.6 | +2.8 -2.8 | +1.3 -1.0 | +3.2 -3.1 | +2.0 -0.9 |

**Table 6 – Performance comparison of road and trip pricing schemes vs. corresponding revenue-neutral pricing schemes, designed with a single objective (traffic efficiency) or with all objectives. For the interpretation of symbols and colours, please refer to Table 5.**

Revenue-neutral schemes confirmed their ability to improve users' acceptance of travel alternatives with respect to the reference scenario, even when maximizing traffic efficiency only (see -11% and -17% for revenue-neutral road and trip pricing schemes, respectively). Results from traffic efficiency optimization suggest that, *in trip pricing*, the compensation between positive and negative prices through revenue-neutral constraints allows reducing congestion (almost) as much as without constraints, though with lower users' perceived costs, i.e., higher users' acceptance, at no expenses for the network regulator. However, in case of traffic efficiency optimization, the detrimental effect on spatial equity produced by revenue-neutral schemes is even worse than in case of pricing schemes without such constraints (+132% vs. +740% for road pricing and revenue-neutral schemes, and +70% vs. +102% for trip pricing and revenue-neutral schemes). This effect is explained by the contrasting nature of spatial equity and traffic efficiency objectives (see the discussion in Section 4.3.2).

Clearly, such detrimental effect on spatial equity could not be appreciated in case of multi-objective optimization. Therefore, these results revealed that the adoption of a multi-objective framework to design revenue-neutral schemes is crucial to preserve spatial equity. However, it is worth noting that the relative improvement of traffic efficiency with respect to the reference scenario is much lower in multi-objective revenue-neutral schemes than in pricing schemes. In fact, incentives provided in in a car-only network allows increasing spatial equity by shifting cars towards less attractive/longer paths, at the expenses of traffic efficiency.

## 4.4. Results in the multimodal network

### 4.4.1. Reference scenario

As anticipated in Section 4.2, the demand values assigned in the multimodal network have been set to values that return SUE car flows (without pricing) equal to the car demand flow applied in the car-only network scenario (Section 4.3.1).

Figure 3 depicts the multimodal SUE link flow distribution in the scenario without pricing. Basing on the above consideration, the SUE flow distribution on car network coincides with the flow distribution in the car-only network scenario shown in Figure 2. In the multimodal network, e-Bikes and metro passenger flows amount, on average, to the 28% of the total demand flow. The distribution of modal shares for the four OD pairs are shown in Figure 4 showing different levels of availability to other modes for different OD pairs as it may be the case for areas with or without access to efficient transit services.

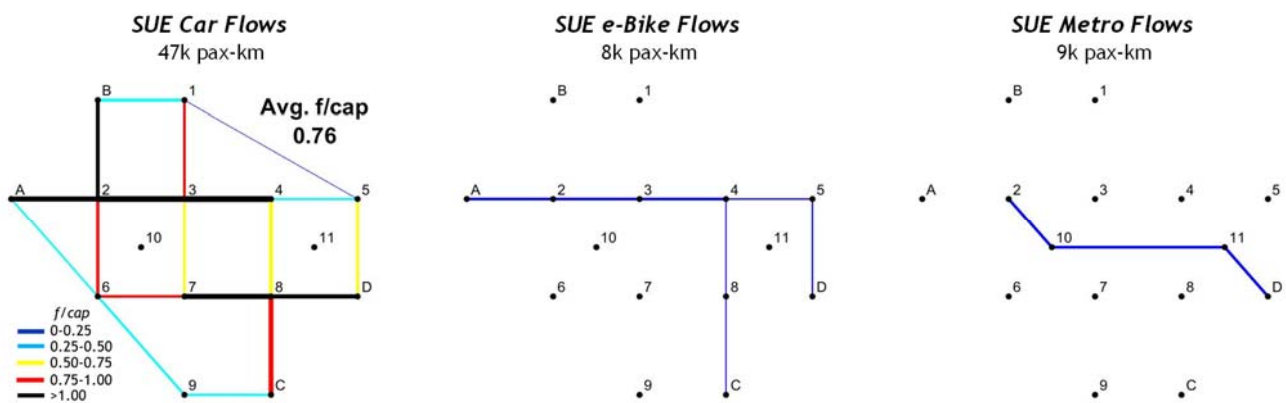

**Figure 3 – SUE flow distribution and passenger traffic in the reference scenario, i.e., multimodal multiclass SUE without pricing. Car network links are coloured according to the flow/capacity ratio.**

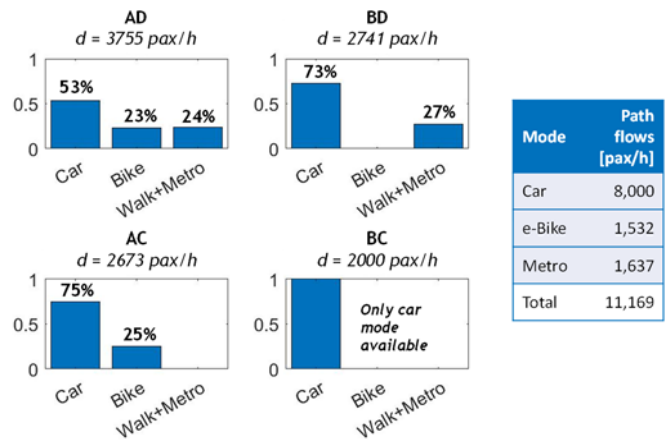

**Figure 4 – Modal split in the reference scenario. OD demand flow values are set in order to achieve a car path flow equal to 2,000 pax/h (network flow on car paths equal to 8,000 pax/h).**

Overall, the multimodal reference scenario depicts a congested road network with limited modal split on more sustainable modes. The total travel time spent in the system is 4.66 h (25 min/pax, on average).

*4.4.2. Monomodal road vs. trip pricing schemes*

Results from second-best pricing in which pricing was restricted only to car network elements, i.e., monomodal pricing, are reported in Table 7.

Results in Table 7 confirm the main findings discussed in Section 4.3.2 in the car-only network scenario. In addition, the negative impact of second-best pricing on users' acceptance is amplified by demand elasticity (compare results with the ones from the car-only network scenario in Table 5). Especially for trip pricing – which was proved to yield smaller prices than road pricing – higher prices than in the car-only network scenario were required to re-balance traffic efficiency and users' acceptance/equity (see 69k€ revenues in the last column of Table 7 vs. 29k€ shown in the corresponding column of Table 5).

It is worth noting that the revenues for the pricing operators, either in road or in trip pricing schemes, are higher than the revenues for the highway and metro operators, no matter if single-objective or multi-objective optimization is carried out to design the pricing scheme. Therefore, if the pricing scheme were operated by a public agency, highway tolls and subway fares could be fully subsidized by the pricing operator.

Focusing on the modal split of more sustainable modes, the highest values are observed in the single-objective optimization targeting at environmental goals (on average, 74% in road and trip pricing, compared to 28% in the reference scenario). Clearly, such benefit is paid at the expenses of traffic efficiency. Overall, as for the car-only network scenario in Section 4.3.2, also in a multimodal network, the "best-compromise" monomodal (second-best) pricing solution is obtained through a multi-objective optimization approach (-42% and -25% for road and trip pricing, on average among all MoPs, with respect to the reference scenario; see the 8[th] and last column in Table 7).

| MoP | Ref. | Monomodal Road pricing (car) Objective function | | | | | | Monomodal Trip pricing (car) Objective function | | | | | |
|---|---|---|---|---|---|---|---|---|---|---|---|---|---|
| | | *Eff.* | *Env.* | *Acpt.* | *Soc. Equ.* | *Spat. Equ.* | *All* | *Eff.* | *Env.* | *Acpt.* | *Soc. Equ.* | *Spat. Equ.* | *All* |
| ΔTTS | | -48% | -19% | - | -21% | -7% | -28% | -55% | -19% | -46% | -21% | -28% | -33% |
| ΔTEC | | -19% | -70% | - | -70% | -30% | -55% | -8% | -71% | -4% | -69% | -42% | -48% |
| ΔPC | | +37% | +431% | - | +419% | +89% | +108% | -1% | +333% | -3% | +413% | +137% | +36% |
| ΔMAPD$_Q$ | | -19% | -74% | - | -75% | -42% | -64% | -12% | -68% | -3% | -75% | -56% | -62% |
| ΔMAPD$_W$ | | -15% | -68% | - | -65% | -90% | -87% | -15% | -8% | -3% | -65% | -100% | -100% |
| *Average* | | *-13%* | *+40%* | *-* | *+38%* | *-16%* | *-25%* | *-18%* | *+33%* | *-12%* | *+36%* | *-18%* | *-42%* |
| Avg. Travel Time [min/pax] | 25 | 13 (-48%) | 20 (-19%) | - | 20 (-21%) | 23 (-7%) | 18 (-28%) | 11 (-55%) | 20 (-19%) | 14 (-46%) | 20 (-21%) | 18 (-28%) | 17 (-33%) |
| Car traffic [kpax-km] | 47 | 46 (-2%) | 12 (-74%) | - | 13 (-72%) | 31 (-34%) | 21 (-55%) | 55 (+17%) | 12 (-74%) | 52 (+10%) | 13 (-72%) | 26 (-45%) | 26 (-45%) |
| Avg. f/cap ratio | 0.76 | 0.54 (-29%) | 0.23 (-70%) | - | 0.23 (-70%) | 0.59 (-22%) | 0.36 (-52%) | 0.55 (-28%) | 0.20 (-74%) | 0.65 (-14%) | 0.23 (-70%) | 0.49 (-36%) | 0.40 (-47%) |
| e-Bike & metro traffic [kpax-km] | 17 | 22 (+29%) | 48 (+182%) | - | 48 (+182%) | 20 (+18%) | 39 (+129%) | 16 (-1%) | 48 (+182%) | 16 (-1%) | 48 (+182%) | 34 (+100%) | 36 (+112%) |
| e-Bike & metro split | 28% | 32% (+14%) | 74% (+164%) | - | 73% (+161%) | 45% (+61%) | 61% (+118%) | 25% (-11%) | 74% (+164%) | 24% (-14%) | 73% (+161%) | 54% (+93%) | 56% (+100%) |
| Total revenue from pricing [k€] | - | 50 | 56 | - | 56 | 53 | 71 | 7.4 | 6.3 | 1.7 | 56 | 63 | 69 |
| Total revenue from highway tolls [k€] | 1.3 | 2.0 (+53%) | 0.2 (-84%) | - | 0.3 (-77%) | 0.5 (-62%) | 0.5 (-62%) | 3.0 (+131%) | 0.3 (-77%) | 2.2 (+70%) | 0.3 (-77%) | 0.4 (-69%) | 0.8 (-38%) |
| Total revenue from metro fares [k€] | 3.3 | 4.3 (+30%) | 9.2 (+179%) | - | 9.2 (+179%) | 5.5 (+67%) | 7.6 (+130%) | 3.2 (-1%) | 9.2 (+179%) | 3.2 (-1%) | 9.2 (+179%) | 6.8 (+106%) | 7.1 (+115%) |
| Revenue per pax from pricing [€/pax] | - | 4.5 | 5.0 | - | 5.0 | 4.8 | 6.4 | 0.7 | 0.6 | 0.1 | 5.0 | 5.7 | 6.2 |

**Table 7 – Performance comparison of road and trip pricing schemes applied to the car network and designed with either a single objective or with all objectives. For the interpretation of symbols and colours, please refer to Table 5.**

*4.4.3. Multimodal pricing vs. revenue-neutral schemes*

Table 8 shows the results of the performance comparison between road/trip first-best (multimodal) pricing, i.e., where all network elements can be priced despite the mode type, and second-best (monomodal, i.e., car-only) pricing (taken from Table 7). In addition, the comparison also includes the revenue-neutral variants of first-best road and trip pricing schemes.

| MoP | Ref. | Monomodal Road pricing Obj. function | | Monomodal Trip pricing Obj. function | | Multimodal Road pricing Obj. function | | Multimodal Trip pricing Obj. function | | Revenue-neutral Multimodal Road pricing Obj. function | | Revenue-neutral Multimodal Trip pricing Obj. function | |
|---|---|---|---|---|---|---|---|---|---|---|---|---|---|
| | | Eff. | All | Eff. | All | Eff. | All | Eff. | All | Eff. | All | Eff. | All |
| $\Delta$TTS | | -48% | -28% | -55% | -33% | -57% | -38% | -63% | -33% | -54% | -39% | -60% | -30% |
| $\Delta$TEC | | -19% | -55% | -8% | -48% | +5% | -23% | +4% | -48% | +1% | -37% | +3% | -58% |
| $\Delta$PC | | +37% | +108% | -1% | +36% | +886% | +174% | +552% | +37% | +46% | +33% | +151% | +12% |
| $\Delta$MAPD$_Q$ | | -19% | -64% | -12% | -62% | -74% | -63% | -74% | -63% | -32% | -33% | -48% | -33% |
| $\Delta$MAPD$_W$. | | -15% | -87% | -15% | -100% | +10% | -65% | -5% | -100% | +96% | -100% | +94% | -100% |
| *Average* | | *-13%* | *-25%* | *-18%* | *-42%* | *+154%* | *-3%* | *+83%* | *-41%* | *+11%* | *-35%* | *+28%* | *-42%* |
| Avg. Travel Time [min/pax] | 25 | 13 (-48%) | 18 (-28%) | 11 (-55%) | 17 (-33%) | 11 (-57%) | 16 (-38%) | 9 (-63%) | 17 (-33%) | 12 (-54%) | 15 (-39%) | 10 (-60%) | 18 (-30%) |
| Car traffic [kpax-km] | 47 | 46 (-2%) | 21 (-55%) | 55 (+17%) | 26 (-45%) | 63 (+34%) | 41 (-13%) | 65 (+38%) | 26 (-45%) | 59 (+26%) | 36 (-23%) | 62 (+32%) | 21 (-56%) |
| Avg. f/cap ratio | 0.76 | 0.54 (-29%) | 0.36 (-52%) | 0.55 (-28%) | 0.40 (-47%) | 0.64 (-16%) | 0.57 (-25%) | 0.60 (-21%) | 0.40 (-47%) | 0.61 (-20%) | 0.44 (-42%) | 0.61 (-20%) | 0.30 (-61%) |
| e-Bike & metro traffic [kpax-km] | 17 | 22 (+29%) | 39 (+129%) | 16 (-1%) | 36 (+112%) | 10 (-41%) | 24 (+41%) | 10 (-41%) | 36 (+112%) | 14 (-18%) | 30 (+76%) | 11 (-35%) | 40 (+135%) |
| e-Bike & metro split | 28% | 32% (+14%) | 61% (+118%) | 25% (-11%) | 56% (+100%) | 13% (-54%) | 33% (+18%) | 13% (-54%) | 56% (+100%) | 17% (-39%) | 31% (+10%) | 15% (-46%) | 73% (+160%) |
| Total net revenue from pricing [k€] | - | 50 | 71 | 7.4 | 69 | 103 | 91 | 19 | 70 | 0.8 | 0.6 | 1.0 | 0.6 |
| Total revenue from highway tolls [k€] | 1.3 | 2.0 (+53%) | 0.5 (-62%) | 3.0 (+131%) | 0.8 (-38%) | 3.5 (+169%) | 1.6 (+23%) | 3.9 (+200%) | 0.8 (-38%) | 3.1 (+54%) | 1.7 (+31%) | 3.4 (+161%) | 0.7 (-46%) |
| Total revenue from metro fares [k€] | 3.3 | 4.3 (+30%) | 7.6 (+130%) | 3.2 (-1%) | 7.1 (+115%) | 2.6 (-21%) | 7.1 (+115%) | 0.0 (-100%) | 7.1 (+115%) | 3.9 (+18%) | 10.0 (+203%) | 0.6 (-82%) | 6.9 (+109%) |
| Net revenue per pax from pricing [€/pax] | - | 4.5 | 6.4 | 0.7 | 6.2 | 9.2 | 8.2 | 1.7 | 6.3 | +4.7 -5.9 | +9.8 -6.3 | +6.3 -6.9 | +8.3 -5.4 |

**Table 8 – Performance comparison of road and trip pricing schemes vs. corresponding revenue-neutral pricing schemes, designed with either a single objective (traffic efficiency) or with all objectives. For the interpretation of symbols and colours, please refer to Table 5.**

Results show that allowing car-alternative modes to be priced in multimodal (first-best) pricing deteriorates performances, on average, relative to (car-only) monomodal (second-best) pricing (compare columns of monomodal and multimodal pricing, for both road and trip pricing). Such negative effect is mostly due to the decrease of users' acceptance and spatial equity, which is not compensated by the increased traffic efficiency. Traffic efficiency optimization further amplifies this effect, since pricing alternative modes diminishes travel times by pushing users towards car paths, at the expenses of increased consumptions, perceived costs, and spatial inequity. In case of multi-objective optimization, instead, a more balanced solution among all objectives is reached.

Introducing revenue-neutral constraints in multi-objective trip pricing optimization, a solution yielding the best-compromise among all objectives is achieved (see the last column in Table 8).

The performances of such solution are comparable with the ones achieved in trip pricing in the car-only network, which is a benchmark to compare results on multimodal network (compare the 6[th] and last columns in Table 8).

Concerning the modal split to sustainable modes, the best-compromise solution resulting from revenue-neutral multimodal trip pricing designed in a multi-objective optimization framework yielded the higher percentage of demand flow assigned to e-Bikes and metro paths. In turn, the general reduction of car path flows with respect to reference scenario, also affects the highway traffic and produces a reduction of total revenues from highway tolls. Such effect may thus require alternative policy instruments to cope with it.

## 5.   SOLUTIONS FOR PRACTICAL IMPLEMENTATION OF TRIP PRICING MODELS

To implement trip pricing schemes, the system operator should be able to distinguish paths of individual users. In the recent literature, the adoption of automated vehicle identification (AVI) sensors, located on specific links that allow an increasing observability of paths, was investigated (Zangui et al., 2015b). Such approach grounds on consolidated methods developed in demand modelling to address the link count sensor location problem in OD flow estimation/correction, though suffering their same limits. In fact, it is well-acknowledged that the performance of path flow estimation using link observations is highly influenced by the unbalance between the information provided by links and the number of path flows to estimate. Reducing the dimensionality of the problem, i.e., the number of paths to monitor, is thus key to effectively observe path flows from link counts (e.g., Marzano et al., 2009). Conversely, if one reduces the space of priceable paths to increase path observability through AVI sensors, congestion pricing scheme performances may deteriorate.

Moreover, the adoption of AVI sensors would inherently limit the applicability of path-based pricing schemes to car users only.

To fully exploit the potential of trip pricing schemes, in this work we propose to use a smartphone application which allows individual tracking of users' paths via GPS, at a level of detail which is fully compliant with data privacy regulations such as the EU GDPR. To incentivize app downloads and increase adoption, trip pricing schemes could be coupled with cordon-based schemes designed with (high) tolls to be charged only to non-app users.

The implementation of revenue-neutral variants of trip pricing schemes would also benefit from such app-based system, via users' in-app wallets that can be charged or credited.

Eventually, the collection of users' paths through the app would also allow the system operator to limit the design of priceable path to the selection of actual users-followed paths, solving computational complexity issues discussed in Section 3.1.

In turn, a so-designed trip pricing scheme would require to explicitly account for the privacy cost in the users' acceptance function (Ren et al., 2020). Following the top-down approach discussed in Section 3.2.3., the inclusion of such additional cost in the users' perceived generalized cost of travel – which contributes to the satisfaction function of path costs (see Eq. 17) – is straightforward.

## 6. CONCLUSIONS

Basing on a thorough review of the literature on pricing schemes for congestion management, we performed a systematic comparison of the performances of price-based models once designed either with a single objective (traffic efficiency, environment, users' acceptance, social equity, or spatial equity, alternatively) or with all multiple objectives. It is believed that the potential to reach social objectives through price-based policies regulating travel demand has been largely unexpressed in real-world applications also because of the way such policies have been conceived and communicated.

To this aim, a comprehensive modelling framework, which extends and generalize available formulations proposed so far in the literature to design congestion pricing schemes, was here introduced to fairly compare the performances of alternative models against multiple pricing objectives. It is worth noting that the proposed modelling framework does not include tradable credit-based schemes since they require the establishment of an external market to buy and sell mobility credits, which is beyond the scope of this comparative analysis.

The proposed approach – here referred to as *trip pricing* model – extends design approaches under a multiclass and multimodal *stochastic* user equilibrium assignment model. In addition, the model takes into account revenue-neutral constraints at the path level, *without* formulating any a-priori assumption on what paths/modes to be tolled/incentivized. Through the proposed framework, it was shown that the traditional link-based pricing model, i.e., the *road pricing* model, is a special case of the *trip pricing* model. In particular, link prices in road pricing were proven to be a source of additive path costs, thus making the degree of freedom of a road pricing system always lower than the one of a trip pricing system. Through numerical experiments, this property was shown to have profound effects on road pricing performances, compared to trip pricing ones.

We formulated the trip pricing design problem under multiple objectives, by combining *five main pricing objectives* in a scalarized objective function, i.e., traffic efficiency, environment sustainability, users' acceptance, social equity, and spatial equity. In particular, two novel measures of performance, based on the satisfaction variable among travel alternatives, were proposed to quantify users' acceptance and social/spatial equity.

Numerical experiments were performed to compare performances of road and trip pricing models once designed either with a single objective (traffic efficiency, environment, users' acceptance, social equity, or spatial equity, alternatively) or with all multiple objectives. Eventually, a comparison between pricing schemes and their revenue-neutral variants was carried out. Clearly, the absolute value of models' performances measured in the numerical exercise were not intended to be generally transferable to different network configurations. Rather, we aimed at identifying the range of variability of pricing scheme performances among different design configurations.

The experiments were carried out on the Nguyen-Dupuis network. Pricing performances were evaluated both in a car-only network scenario composed by car links with a fixed demand, and in a multimodal network scenario with elastic demand on mode choice over car, e-Bike and metro path alternatives.

In the car-only network scenario, results revealed that in trip pricing highly priced paths always return low path flows, in contrast to road pricing. Consequently, path flows are less dispersed in trip pricing than in road pricing, no matter high prices. Trip pricing was found to be always more efficient, environmentally sustainable, acceptable, and equitable than road pricing. In addition, at the path-level, the total path price paid by users in road pricing was always greater than the one paid in trip pricing. This result was due to the additive nature of link prices in road pricing, which yields a system with less degrees of freedom than the trip pricing one. As expected, user's acceptance revealed to be the most challenging measure of performance to optimize. In fact, higher prices which are set to redistribute congestion in space, i.e., to encourage the choice of longer paths, are not compensated by the corresponding travel time savings, thus deteriorating users' satisfaction. This effect called for revenue-neutral schemes, which are able to maximize users' acceptance through incentives. Moreover, spatial equity was found harder to increase than social equity, being structurally dependent on the topology of the network rather than on the transportation system performances (which can be controlled by pricing). In addition, spatial equity and traffic efficiency were shown to be contrasting objectives. In fact, if spatial equity is maximized, shifting users with alternative paths from shortest to longer/less attractive paths, deteriorates the total time spent in the system.

Evidence from the application of the proposed multi-objective design framework revealed that the trip pricing model performances improves significantly on all objectives, with respect to a single-objective design.

In the multimodal network scenario, results showed demand elasticity makes users' acceptance even harder to improve (than in the car-only network experiment). For instance, in case of multi-objective optimization, second-best trip pricing scheme – which was proved to yield smaller prices than road pricing ones – required higher prices (than the ones applied in the car-only network scenario) to re-balance traffic efficiency and users' acceptance.

Even if all paths were allowed to be priced (i.e., first-best pricing), revenues for the pricing operators, both in road and trip pricing schemes, were found to be higher than the revenues for the highway and metro operators, no matter if single-objective or multi-objective design was carried out. Therefore, if the pricing scheme were operated by a public agency, highway tolls and subway fares could be fully subsidized by the pricing operator.

Eventually, the introduction of revenue-neutral constraints in the multi-objective trip pricing on the multimodal network, yielded the best-compromise solution over all five objectives, with performances, on average, higher than the ones obtained by road and trip pricing models without such constraints.

Numerical results suggest a vast potential for trip congestion pricing as opposed to road pricing alone. Also, the possibility of applying both positive and negative (incentives) prices was shown to significantly increase the expected acceptance of pricing schemes. Finally, pricing schemes applied in multimodal networks where more energy efficient modes (than car) are available, such as transit and e-bikes, allows reaching significant beneficial effects also for the environment.

The research proposed in this work could be extended over several directions. From a modelling standpoint, the formulation of (bottom-up) pricing acceptance measures based on SP surveys eliciting users' preferences for different pricing models and related consequences, is critical. The model could be also enhanced by taking into account *i)* user class-specific (non-anonymous) prices, *ii)* within-day dynamic traffic networks with time-varying prices and pricing effects on departure time choices. Eventually, the extension of the comparative analysis among different pricing models to real-world transportation networks, would allow extending the validity of the preliminarily results discussed in this work.

## ACKNOWLEDGEMENTS

The authors acknowledge the precious discussions with Prof. Armando Cartenì and Dr. Ilaria Henke. Research in this paper has been partially funded by the Italian program PON AIM - Attraction and International Mobility, Linea 1 (AIM1849341-2, CUP E61G18000540007) and by the Italian program MIUR – Departments of Excellence.

## APPENDIX A

Differences between UE and SO conditions can be easily shown on the two-link network in Figure A.1, taken from Cascetta (2009), in which link flows and path flows are equivalent, i.e., $f = h$, and path costs are purely additive.

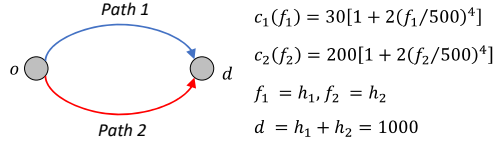

**Figure A.1 – Two-link network and supply/demand models.**

Figure A.2 shows on the x-axis the set of all feasible path flow vectors, i.e., H, that can be loaded on the two paths. Blue and red solid curves depict the cost functions of path 1 and path 2, respectively. The deterministic UE flow distribution $f^{UE}$ result in equal costs on the two (used) paths, i.e., any flow unit shifted from one path to the other would cause an increase of the user cost on the other path, which would violate Eq. 2. Conversely, blue and red dashed line depict the total cost functions of the two paths, i.e., $c_i(f_i)f_i = g_i(h_i)h_i$, with $i = 1, 2$. The grey dashed line shows the total cost function at the network level, i.e., $c(f)^T f = g(h)^T h$, whose minimum point, $f^{SO}$, corresponds to the SO flows (see Eq. 3).

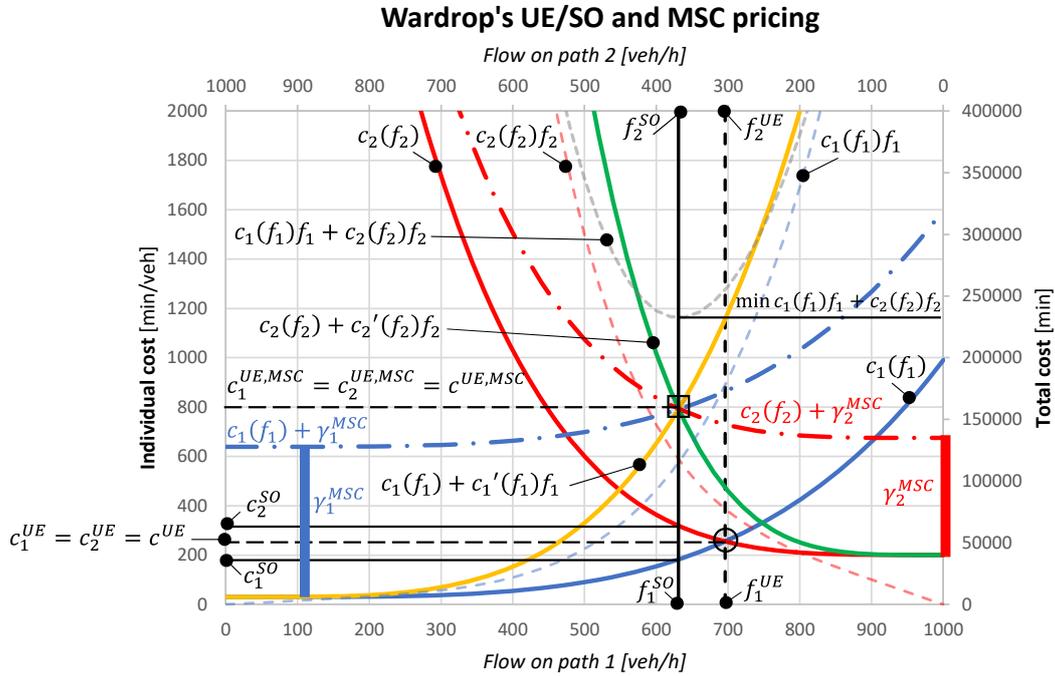

**Figure A.2 – Wardrop's UE flows (vertical black dashed line) and SO flows (vertical black solid line), and corresponding individual generalized costs (blue and red solid lines for path 1 and 2 respectively) and total generalized costs (blue and red dashed lines for path 1 and 2, respectively). Green and yellow lines depict the marginal external cost functions for path 1 and path 2, respectively, while blue and red dash-dot lines depict the corresponding individual generalized cost functions with MSC pricing. The circle and the square markers correspond to the UE conditions without pricing and with MSC pricing, respectively.**

If MSC tolls $\pi^{MSC} = g^{ad}(h^{SO})'h^{SO} = c(f^{SO})'f^{SO} = \gamma^{MSC}$ are added to path/link cost functions, the corresponding cost functions are graphically translated up in Figure 2, so that the costs on both paths/links become equal, i.e., UE with MSC pricing yields a SO flow distribution (see blue and red dash-dot lines in Figure A.2).

However, MSC pricing allows reaching the SO condition only under the assumption of deterministic traffic assignment. In case of stochastic assignment, instead, the toll values which guarantee a SO flow distribution depends on the path choice model. Even for the elementary example in Figure A.2, if designed MSC prices, i.e., $\gamma_1^{MSC} = 609$ and $\gamma_2^{MSC} = 475$, were applied in a stochastic assignment based on a logit route choice model with a high dispersion parameter (e.g., equal to 35), the UE flows with MSC will be 2% off than stochastic SO flows.

Eventually, MSC tolls, i.e., $\gamma^{MSC}$, do not correspond to the minimum toll allowing to reach the SO condition, therefore hindering the users' acceptance of such pricing scheme. Figure A.3 shows alternative valid toll vectors, which are obtained by designing $\gamma$ such that the UE condition is reached at SO (see the dash-dot blue and red lines).

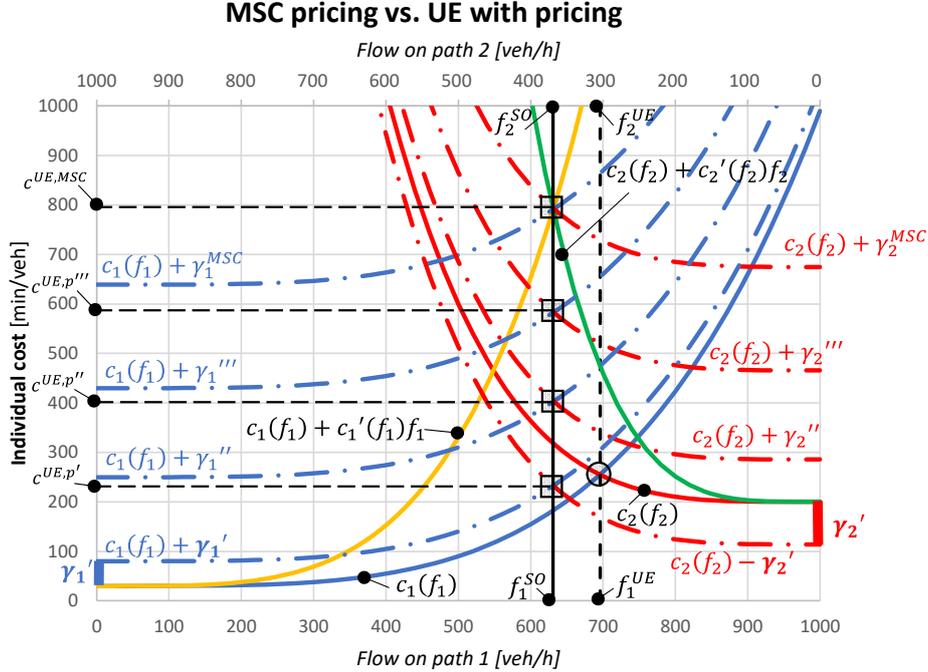

**Figure A.3 – MSC pricing interpreted as the result of a non-additive path cost summed to the individual generalized path cost function. Dash-dot blue and red lines show alternative price combinations yielding SO flows at equilibrium, corresponding to lower individual generalized costs at equilibrium. In particular, the price combination ($\gamma'_1, \gamma'_2$) admits both positive prices ($\gamma'_1$) and negative prices ($\gamma'_2$), i.e., incentives, and produces a zero net revenue.**

In the figure, three valid price vectors, $\gamma'$, $\gamma''$ and $\gamma'''$, yielding lower tolls than $\gamma^{MSC}$ are shown. In particular, $\gamma'$ is composed of a positive price (toll) on path 1, and a negative price (incentive) on path 2, producing *i)* a lower generalized user cost than UE cost on both paths (Pareto-improving pricing), and *ii)* a zero net revenue for the pricing operator (revenue-neutral pricing).

## APPENDIX B

In this section the multimodal multiclass stochastic user equilibrium assignment model applied in this work is presented.

The transportation supply is simulated with a congested network model. For each users' class $q$, the relation between path costs, $g^q$, and link costs, $c^q$, is given by:

$$g^{ad,q,m,w} = \Delta_{m,w}{}^T c^q(f) \tag{B.1}$$

$$g^{q,m,w} = g^{ad,q,m,w} + g^{nad,q,m,w} \tag{B.2}$$

In Eq. B.1, the vehicle link flows are a function of vehicle path flows, which is expressed by:

$$f = \sum_{q \in Q} \sum_{m \in M} \sum_{w \in W} \Delta_{m,w} h^{q,m,w} = \sum_{q \in Q} \Delta h^q \tag{B.3}$$

The generalized link cost function $c^q(\cdot)$ depicts the general congestion model.

The whole supply model is defined by Eqs. B.1-3, which are combined to express the relationship between path costs and path flows:

$$g^{q,m,w} = \Delta_{m,w}{}^T c^q \left( \sum_{q \in Q} \sum_{m \in M} \sum_{w \in W} \Delta_{m,w} h^{q,m,w} \right) + g^{nad,q,m,w} \tag{B.4}$$

The transportation demand is formulated as a RU model. As discussed in Section 3.1, demand flows are known and are independent of cost variations. It is also assumed that the demand flows are expressed in passenger units.

Mode and path choices are modelled through a single-level hierarchical logit model, with dispersion parameters $\theta^M$ and $\theta^K$. Linear systematic utility functions are assumed for mode and path choice alternatives.

Given a users' class $q$ and OD pair $w$, the mode choice probability is a function of the systematic utility of mode $m$, $V_m^{M|q,w}$. Such utility is assumed equal to the Expected Maximum Perceived Utility (EMPU; Cascetta, 2009) among path alternatives $K_{w,m}$, i.e., $s\left(V^{K_{m,w}|q}\right)$. The EMPU variable is also referred to as *satisfaction* in the field literature (Ben-Akiva and Lerman, 1975):

$$V_m^{M|q,w} = s\left(V^{K_{m,w}|q}\right), \ \forall m \in M : \left|K_{m,w}\right| > 0 \tag{B.5}$$

where $V_k^{K_{m,w}|q}$ is the systematic utility of path $k$, given by:

$$V_k^{K_{m,w}|q} = -g_k^{q,m,w} \tag{B.6}$$

The choice probability for users of class $q$ of mode $m$ between OD pair $w$ is a function of $V^{M|q,w}$:

$$p_m^{M|q,w} = p_{mode}\left(V^{M|q,w}\right), \forall m \in M : \left|K_{m,w}\right| > 0 \tag{B.7}$$

The choice probability for users of class $q$ of path $k$ between OD pair $w$, conditional to the choice of mode $m$, is a function of $V^{K_{m,w}|q}$:

$$p_k^{K_{m,w}|q} = p_{path}\left(V^{K_{m,w}|q}\right), \forall k \in K_{m,w} \tag{B.8}$$

Therefore, for users of class $q$, given the passenger demand flow, the vehicle path flow vector on paths of mode $m$ between OD pair $w$ is given by:

$$h^{q,m,w} = p^{K_{m,w}|q} \cdot p^{M|q,w} \cdot \frac{\psi_w^q}{\eta^{q,m}} \cdot d_w \tag{B.9}$$

where $p_m^{M|q,w}$ is the probability for users of class $q$ of choosing mode $m$ between OD pair $w$, and $p_k^{K_{m,w}|q}$ is probability for users of class $q$ of choosing path $k$ between OD pair $w$, conditional to the choice of mode $m$. In Eq. B.9, $\psi_w^q$ is the fraction of total demand flow $d_w$ related to users' class w.

As path flows are expressed in vehicle units (relative to the corresponding mode $m$ and users' class $q$), passenger demand flows in Eq. B.9 are converted to mode- and users' class-specific vehicle flows via the occupancy factor $\eta^{q,m}$.

The multimodal multiclass fixed-point user equilibrium model results from combination of Eqs. B.3, B.4, B.6, B.7, B.8 and B.9:

$$f^{SUE} = \sum_{\substack{q \in Q \\ m \in M \\ w \in W}} \Delta_{m,w} p^{K_{m,w}|q} \left( -\Delta_{m,w}{}^T c^q(f^{SUE}) - g^{nad,q,m,w} \right) p^{M|q,w} \left( -\Delta_{m,w}{}^T c^q(f^{SUE}) - g^{nad,q,m,w} \right) \frac{\psi_w^q}{\eta^{q,m}} \cdot d_w \tag{B.10}$$

For the sake of readability, in the remaining of the paper, Eq. B.10 is expressed as:

$$f^{SUE} = \sum_{\substack{q \in Q \\ m \in M \\ w \in W}} \Delta_{m,w} p^{q,m,w} \left( -\Delta_{m,w}{}^T c^q(f^{SUE}) - g^{nad,q,m,w} \right) \frac{\psi_w^q}{\eta^{q,m}} \cdot d_w \tag{B.11}$$

where the path choice probability vector $p^{q,m,w}$, for users of class $q$, is given by:

$$p^{q,m,w} = p_k^{K_{m,w}|q} \cdot p_m^{M|q,w}, \forall w \in W \tag{B.12}$$